\documentclass[preprint,amsmath,superscriptaddress,prl]{revtex4}  
\usepackage[dvips]{graphicx}
\usepackage{epsfig}
\usepackage{color}
\usepackage{subfigure}
\usepackage{soul}

\begin{document} 

\title{
The distinct flavors of Zipf's law
in the rank-size 
and in the size-distribution
representations, 
and its maximum-likelihood fitting
}
\author{\'Alvaro Corral}
\affiliation{%
Centre de Recerca Matem\`atica,
Edifici C, Campus Bellaterra,
E-08193 Barcelona, Spain
}\affiliation{Departament de Matem\`atiques,
Facultat de Ci\`encies,
Universitat Aut\`onoma de Barcelona,
E-08193 Barcelona, Spain
}\affiliation{Barcelona Graduate School of Mathematics, 
Edifici C, Campus Bellaterra,
E-08193 Barcelona, Spain
}\affiliation{Complexity Science Hub Vienna,
Josefst\"adter Stra$\beta$e 39,
1080 Vienna,
Austria
}
\author{Isabel Serra}
\affiliation{%
Centre de Recerca Matem\`atica,
Edifici C, Campus Bellaterra,
E-08193 Barcelona, Spain
}
\author{Ramon Ferrer-i-Cancho}
\affiliation{Complexity and Quantitative Linguistics Lab, 
Departament de Ci\`encies de la Computaci\'o, Universitat Polit\`ecnica de Catalunya, Barcelona, Catalonia, Spain}

\begin{abstract} 
In the last years, 
researchers have realized the difficulties of fitting power-law distributions properly.
These difficulties are higher in Zipf's systems, 
due to the discreteness of the variables and
to the existence of two representations for these systems, i.e.,
two versions about which is the random variable to fit.
The discreteness implies that a power law in one of the representations
is not a power law in the other, and vice versa.
We generate synthetic power laws in both representations
and apply a state-of-the-art fitting method 
(based on maximum-likelihood plus a goodness-of-fit test)
for each of the two random variables.
It is important to stress that the method does not fit the whole distribution, 
but the tail, understood as the part of a distribution above a cut-off
that separates non-power-law behavior from power-law behavior.
We find that, no matter which random variable is power-law distributed,
the rank-size representation is not adequate for fitting, whereas 
the representation in terms of the distribution of sizes leads to 
the recovery of the simulated exponents, may be with some bias.

Keywords:
Statistical inference;
Scaling in socio-economic systems;
Zipf's law.
\end{abstract} \maketitle

\section{Introduction}

Power-law distributions are supposed to show up in the statistics of many complex systems
\cite{Bak_book,Sornette_critical_book,Mitz,Newman_05,Simkin11}.
However, the fitting of power-law distributions, 
or more precisely, power-law tails, is a delicated issue
\cite{Bauke, White,Clauset,Corral_nuclear,Corral_Deluca,Corral_Gonzalez,Voitalov_krioukov}.
Zipf's systems constitute a very special but important area in complex-systems science
for which power-law fitting is particularly involving.
Since its formulation, more than 100 years ago, 
Zipf's law has emerged
as a paradigm in the statistics of social, economic, cognitive, biological, and technological processes.
Indeed, 
this law, of empirical nature, has been proposed to hold for many different
complex systems 
{\cite{Zipf1949a,Newman_05,Adamic_Huberman,Furusawa2003,Axtell,Serra_scirep}}.

Zipf's systems can
be described in the following simple way. 
Let us consider a system composed by some entities,
which we call types,
and that each of these types can be characterized by a certain discrete ``size'';
further, each elementary unit of size will constitute a token.
If the system is a text, 
each appearance of a word is a token
corresponding to the type given by the word itself;
then, the size of a type will be its number of appearances
(i.e., its absolute frequency) \cite{
Baayen,Baroni2009,Piantadosi}.
For a country, the tokens will be its citizens, whereas the types can be the
cities where they live \cite{Malevergne_Sornette_umpu}, 
or their family names \cite{}, 
etc.;
the size of each type will be the population associated to it.
Another possibility is that the tokens are monetary units 
(let us say, richness translated into one-dollar pieces) 
and the types are the persons owing that money;
the measure of the ``size'' of one person will be, in these terms, his/her richness
(for other examples see, e.g., Ref. \cite{Lu_2010} and Table \ref{tablezero}).

Zipf's law deals with how these tokens are distributed into types.
%
%
%
%
%
Counting the number of tokens that
correspond to any type allows one not only to define 
the size $n$ of the types but also their rank $r$, which is the position of 
each type when these are ordered by decreasing size. 
{Then, $n(r)$ is the number of tokens of the type with rank $r$.} 
{For instance, in the book {\it Ulysses} (by James Joyce),
the rank $r=1$ corresponds to type {\it the}, as this is the most common word type;
for the population of US cities, $r=1$ corresponds to New York; 
and for US wealth, the person ranked at the top is William Henry Gates III;
the size of these types is $n(r=1)=14,934$ appearances, 
8,622,698 inhabitants, and
$95 \times 10^9$ dollars, 
respectively 
(at the moment of writing this article).}
If several types have the same size
{(which is common for low sizes)},
the assignation of the rank among them is arbitrary.

The dependence between size $n$ and rank $r$ 
(necessarily non-increasing)
yields the rank-size (or size-rank) or rank-frequency relation, and
a first form of
Zipf's law holds when 
both $n$ and $r$ are related through a decreasing power law, i.e.,
\begin{equation}
n(r) \propto \frac 1 {r^\alpha},
\label{laprimera}
\end{equation} 
with exponent 
$\alpha$ more or less close to one,
and the symbol ``$\propto$'' denoting proportionality.
This formulation will be referred to as the Zipf's law for types, 
as it is obtained from the statistics of types
(counting their repetitions, i.e., their tokens).

\begin{table}[h]
\begin{center}
\caption{\label{tablezero} 
Diverse examples of Zipf's systems, 
together with their corresponding tokens and types.
Note that we do not sustain that a Zipf's system
has to fulfill Zipf's law (in any of the two forms considered in this article).
RW denotes a random walk.
}
\smallskip
\begin{tabular}{|lll|}
\hline
System/discipline & tokens & types\\
\hline
Texts & occurrences of words & words ``themselves''  \\
Songs & occurrence of chords & chords ``themselves'' \\
Bibliometry & citations & papers \\
RW in networks & visits & sites\\
\hline
Cells & counts of molecules & molecules themselves \\
\hline
Beliefs & individuals & religions\\
Demography & individuals & cities\\
Demography & individuals & family names\\
Bibliometry & papers & authors \\
Ecology & individual insects & individual plants\\
Economics & employees & companies \\
Economics & pieces of \$1 & individuals \\
\hline
Networks & links & nodes\\
Telephony & calls & people\\
Internet & connections & computers \\
\hline
RW in networks & occurrences of transitions& transitions themselves \\
Texts & occurrences of bigrams & bigrams themselves \\
\hline
\end{tabular}
\par
\end{center}
\end{table}


An equivalent description of this sort of systems is possible in terms of the distribution of sizes
(or distribution of frequencies). 
For that, one counts not only the repetitions 
of each type (i.e., 
its size $n$) 
but also the repetitions of each size, 
i.e., one counts the number of types with a given size \cite{Baroni2009,Zanette2012a}.
In probabilistic terms this means that the size {of the types} is
considered as a random variable, which is for which the statistics is performed.
Then, $f(n)$, the probability mass function of $n$,
is the quantity of interest.
Zipf's law for sizes holds when
\begin{equation}
f(n) \propto \frac 1 {n^\gamma},
\label{pretercera}
\end{equation}
where, remember, the symbol ``$\propto$'' denotes proportionality
(not asymptotic behavior).
Many authors have argued or assumed that both forms of Zipf's law, 
Eq. (\ref{laprimera}) and Eq. (\ref{pretercera}), are equivalent \cite{Adamic_Huberman,Newman_05,Zanette2012a},
with a relation between their exponents given by 
\begin{equation}
\gamma=1+ \frac 1 \alpha,
\label{relationship_between_exponent_equation}
\end{equation}
but the equivalence only holds exactly asymptotically, for large $n$,
i.e., for low ranks
\cite{Mandelbrot61}. We will turn to this important issue below.

{
In the second version of the law 
it may seem strange that
one needs to perform double statistics -- 
{
first the statistics of types, counting tokens to obtain 
the size of every rank (i.e., of every type),
and then the statistics of sizes, counting types, 
to obtain the number of types of a given size.
This is indeed the case in text statistics, where the frequency
obtained counting tokens takes the role of the random variable, 
which needs to be counted also, within this framework.
In contrast, 
in other systems where Zipf's law is supposed to hold, 
such as} cities, 
Zipf's law
can be obtained directly for the statistics of
the 
sizes $n$ of the studied entities, 
which are usually known from the beginning
{(we do not need to go city by city counting
all their inhabitants).
Nevertheless, this does not constitute a fundamental difference
between both kinds of systems,
and the only difference comes in the way the data are usually available.}
}

In order to dissipate any misunderstanding,
it is useful to clarify what $n(r)$ and $f(n)$ mean in practice.
If one picks randomly a token from a system
(e.g., a person from the census), 
the probability that it corresponds to the type 
(to the city) 
with rank $r$ 
is given by $n(r)/L$
{(where $L$ is system size,
i.e., the total number of tokens, 
i.e., the sum of the sizes of all types)}.
Knowledge of $n(r)$ allows {one} to build a {\it directory} 
(a list of cities, a dictionary...)
with the sizes $n(r)$ of all types;
then, if one picks (randomly) a type from the directory 
(a city from the list of cities), 
the probability that it has size $n$ in the system
is $f(n)$.
{
There is still a third distribution, 
given by $n f(n) / \langle n \rangle$
(if $\langle n \rangle$, the mean of $n$,
is finite), which represents the probability that, 
if one picks randomly a token from system
(a person from the census), it corresponds to any type
(any city) of size $n$.
As this latter distribution is directly related to $f(n)$, 
it will not be considered in this article.
}

%
%
%
%
%
%
%
%
%
%
%
%

In this article we address the question about which the best
approach to verify the fulfillment of Zipf's law is, 
understood either as a power-law relation in the rank-size representation, Eq. (\ref{laprimera}),
or as a power law for the distribution of sizes, Eq. (\ref{pretercera}).
Taking for granted that the most suitable way to fit
power-law (or any other ``well behaved'') probability distribution
\cite{Bauke, White,Clauset}
is maximum likelihood (ML) estimation \cite{pawitan2001},
we will apply this method both to the rank-size relation $n(r)$
and to the distribution of sizes $f(n)$ for simulated systems.
We will use a state-of-the art fitting procedure \cite{Corral_Deluca},
which in addition to ML estimation also incorporates a goodness-of-fit test
(the testing is necessary in order to evaluate the goodness or badness of the ML fit;
ML estimation does not provide goodness of fit).
The fitting procedure pretends to be an improvement of the well-known Clauset et al.'s method
\cite{Clauset}, and it will be applied to power laws without upper truncation; 
these are
power laws that do not have an upper cut-off, i.e., non-truncated power laws
referred here simply as power laws.
(Note that, although Clauset et al.'s method applies only to non-truncated power laws,
the alternative given by Ref. \cite{Corral_Deluca} can be applied to both non-truncated and truncated power laws.)

As the two definitions of Zipf's law (for types and for sizes)
are not fully equivalent, we simulate random systems for the two versions of the law
(and, as mentioned, study each system both using $n(r)$ and $f(n)$).
This yields four different cases of study, 
which are further doubled when one distinguishes
between continuous and discrete distributions.
%
%
In quantitative linguistics,
the overwhelming majority of research has focused on rank as the random variable
{\cite{Baayen,Lu_2010,Gerlach_Altmann,Baixeries2012a}}
(some exceptions are Refs. 
{\cite{Tuldava1996,Balasubrahmanyan1996,Ferrer2000a,Petersen_scirep,Ferrer2009a}),
%
Here 
{we} argue that 
{the alternative track of fitting the distribution of sizes 
has some clear advantages over ranks and is unavoidable 
if one wants to use maximum-likelihood estimation. }
%
%
%
%

Next, we present the statistical frameworks used for modelling Zipf's law 
and how the two representations of this law are not equivalent.
In the third section 
we simulate the two versions of the law and try to recover
the known input parameters using maximum likelihood estimation
applied to the two representations; 
the advantages of maximum-likelihood estimation and its practical application
are briefly explained.
The last section contains a discussion providing further support for
the distribution of sizes, 
and finally, 
an appendix explains the 
complete fitting procedure for the discrete case, including how
to simulate Zipf's law.
This article can be considered a complement or an alternative to Ref. \cite{Altmann_Gerlach}.



%


\section{Differences between Zipf's laws for types and for sizes}

In order to proceed, we need two useful quantities:
the number of tokens $L$, also referred to as size of the 
system,
and
the number of types $V$ (vocabulary for a text).
These are empirical quantities
related through
$$
L = \sum_{r=1}^V n(r).
$$
Of course, $V \le L$, and in any non-trivial case, $V< L$.
It is important to mention that in our analysis
we will not consider data outside the power-law range 
(to be determined below by the fitting procedure),
and therefore, $V$ and $L$ do not correspond 
to the complete empirical data
but to a restricted, truncated data set.
For the complete (total) data set we will use the notation
$V_{tot}$ and $L_{tot}$.

\subsection{Zipf's law for types}

Let us now assume that Zipf's law for types, Eq. (\ref{laprimera}), holds 
empirically
up to some maximum range $r_b$, that is,
\begin{equation}
n(r) = \frac A {r^\alpha}
\label{laprimerabis}
\end{equation}
for $1 \le r \le r_b$, 
with $r_b = V$ 
and with $A$ an appropriate normalization constant
scaling linearly with system size.
(we do not consider the behavior of $n(r)$ for $r> r_b$, 
we assume deviations from the power law we are not interested in).
It turns out to be that Zipf's law for types can be written as
\begin{equation}
S(n) = \frac B {n^\beta},
\label{Sn}
\end{equation}
for $n=n_a, n_a+1\dots$,
with $n_a$ the size associated to $r_b$,
i.e., $n(r_b)=n_a$,
$B$ some constant ($B=A^\beta/V$), 
$\beta=1/\alpha$
and $S(n)$ the 
complementary cumulative distribution of the size,
also called survivor function, {i.e.,
the probability that the size of a type is equal to or above
a particular value $n$; {in a formula,}
$S(n) =$ Prob $[\mbox{size} \ge n]$}.
Indeed, 
by its definition, 
the estimation of $S(n)$ is given by
$S(n) = r/V$
\cite{Newman_05,FontClos_Corral}, 
and if $n$ is not too low the estimation of $S(n)$ 
will be very close to the true $S(n)$;
then, the inversion of Eq. (\ref{laprimerabis}) leads directly to Eq. (\ref{Sn}),
equivalently, see Refs. \cite{Adamic_Huberman,Zanette2012a}.
When several types have the same size,
the $r$ used in the calculation of $S(n)$ 
has to be the one with the largest value among those types
(due to our definition of $S(n)$, which contains the ``$\ge$'' inequality).

{Going one step further,
{we can obtain the probability mass function of $n$
for the Zipf's law for types, as}
$f(n)=S(n)-S(n+1)$ (from the definition $f(n)=$ Prob $[\mbox{size} = n]$);
then, 
$$
f(n) = 
\frac B {n^\beta}\left[1-\left(\frac 1 {1+1/n} \right)^\beta\right]
\simeq
\frac B {n^\beta}\left[1-\left(1-\frac \beta n + \frac {\beta(\beta+1)} {2n^2} +\cdots \right)\right]=
\frac {\beta B} {n^{\beta+1}} \left(1 - \frac {\beta+1} {2n} + \cdots \right)
$$
for $n=n_a,n_a+1,\dots$
(using the binomial theorem and the geometrical series).
A similar result has been obtained for $\beta=1$ in Refs. \cite{Baayen,Heaps_1978}.
So, Zipf's law for tokens, Eq. (\ref{laprimerabis}),
leads to a power-law distribution $f(n) \propto 1 / {n^{\gamma}}$
only for infinitely large $n$, 
with 
\begin{equation}
\gamma=1+\beta=1+\frac 1 \alpha.
\label{gammabetaalpha}
\end{equation} 
We see how a pure discrete power-law form for $n(r)$, 
Eq. (\ref{laprimerabis}), 
does not lead to a pure power law for the probability mass function
of the size, $f(n)$, in the sense that 
although the power law is fulfilled for the rank-size relation,
it will not hold for $f(n)$. 

This issue has received very little attention in the literature, 
as one is usually interested in the fulfillment of Zipf's law
 in an almost qualitative sense, 
for instance just by plotting the logarithm of either 
$n(r)$, $S(n)$, or $f(n)$ versus $\log r$ or $\log n$
and obtaining something reminiscent of a straight line
in some part of the plot
(then, the distinction between a pure and an asymptotic power law
becomes diluted).

A notable exception is provided by Mandelbrot \cite{Mandelbrot61},
who calculated $f(n)$ when 
$L_{tot}$
tokens are drawn randomly and independently of each other
from Zipf's law for tokens, Eq. (\ref{laprimerabis}),
with an infinite population ($r_b \rightarrow \infty$).
In contrast to the previous case, 
Zipf's law in the form of Eq. (\ref{laprimerabis})
is supposed to hold not only for a single empirical sample of the system
but for the underlying population.
Mandelbrot's result, for large $L_{tot}$, was 
\begin{equation}
f(n) 
= \frac{\beta A^\beta 
} V
\frac{\Gamma(n-\beta)}{\Gamma(n+1)}
=\frac{\beta}{\Gamma(1-\beta)}\frac{\Gamma(n-\beta)}{\Gamma(n+1)},
\label{Mandelbrot_Simon}
\end{equation}
for $n=1, 2, \dots$,
with $\Gamma$ the gamma function \cite{Abramowitz}.
For large $n$ the quotient of gamma functions tends to $1/n^{\beta+1}$
(using Stirling's approximation),
and again, one gets a power law for $f(n)$ only asymptotically.
Using the normalization given by the right-hand term 
we find the relation $V=\Gamma(1-\beta) A^\beta$, 
which is essentially Heaps' law \cite{Baayen,FontClos_Corral},
also called the type-token relationship \cite{Mandelbrot61}
(as, as mentioned, $A$ has to scale linearly with system size).
In terms of $S(n)$ one gets
$$
S(n)=\frac{\Gamma(n-\beta)}{\Gamma(1-\beta)\Gamma(n)}.
$$


\subsection{Zipf's law for sizes}

Nevertheless, strictly speaking,
for a random variable $n$ (size)
a discrete power-law distribution is defined in terms
of $f(n)$ and neither in terms of $S(n)$
(although there seems to be some confusion, as in Ref. \cite{Johnson_univariate})
nor in terms of its underlying rank-size relation.
So, when considering the size of types as a discrete random variable,
a power-law distribution would mean that the size probability mass function
is given by
\begin{equation}
f(n) = \frac 1 {\zeta(\gamma,n_a) n^{\gamma}},
\label{fn_sizes}
\end{equation}
for $n=n_a, n_a+1,\dots$
with $\gamma>1$ and 
$\zeta(\gamma,n_a)$ the
Hurwitz zeta function (a generalization of the Riemann zeta function),
defined as
$$
\zeta(\gamma,n_a) = \sum_{k=0}^\infty \frac 1 {(n_a+k)^\gamma},
$$
and providing the normalization of the distribution.
The corresponding cumulative distribution function 
is obtained from
\begin{equation}
S(n) = \sum_{n'=n}^{\infty} f(n'), 
\end{equation}
yielding
\begin{equation}
S(n)=\frac{\zeta(\gamma,n)}{\zeta(\gamma,n_a)},
\label{Sn_sizes}
\end{equation}
which does not have a power-law shape, strictly.
The empirical rank-size relation corresponding to this 
distribution of sizes is given by
\begin{equation}
n(r) = \zeta_2^{-1} \left(\gamma, \zeta(\gamma,n_a) \frac r V \right),
\label{Hurwitz_inversion}
\end{equation}
for $1 \le r \le r_b$,
where $\zeta_2^{-1}$ is the inverse of the Hurwitz zeta function with respect
its second argument.
Note that although, commonly, the empirical frequency takes values in the range $n\ge 1$,
the theoretical $f(n)$ is only defined for $n\ge n_a$, and in general $n_a >1$ or even $n_a \gg 1$. The range $n\ge n_a$ is what we define
as the tail of the distribution, which means that all values of $n$ below $n_a$
are disregarded for the power-law fit (this is within the same philosophy as Clauset et al. \cite{Clauset}).



As outlined in the introduction,
the previous equation for $f(n)$, Eq. (\ref{fn_sizes}),
can be understood as a second, different definition of Zipf's law
(alternative to Eqs. (\ref{laprimerabis})),
which we call Zipf's law for sizes
(as, by counting repetitions of the size variable, it leads to a power law).
Both definitions of Zipf's law 
[Eqs. (\ref{laprimerabis}) and (\ref{fn_sizes})]
are not equivalent, 
only asymptotically equivalent in the limit of large $n$, i.e., small $r$.
The distinction between the definitions 
would not be present if $n$ were a continuous variable,
so, it is an effect of the discreteness of the tokens.
Although the descriptions of systems candidate to fulfill Zipf's law
in terms of the rank-size (or rank-frequency) plot
and in terms of the distribution of sizes are fully equivalent
(in the sense that one can recover any of the two with the knowledge of the other \cite{Baayen,Baroni2009}),
a power-law relationship in one case does not imply  
a power law in the other, and reciprocally.
This leads to the two distinct definitions of Zipf's law
just explained.

In summary, we have two representations of Zipf's systems, in terms of the rank-size
relation or in terms of the distribution of sizes, and both approaches are equivalent
to describe such systems. However, a pure power law in one of the representations
does not imply a pure power law in the other, and vice versa, and therefore we have two alternative,
different definitions of Zipf's law.

\section{Simulations of Zipf's systems and recovering of power-law relationships}

In this section we deal with simulated systems built
using any of the two different versions of Zipf's law discussed above,
Eq. (\ref{laprimera}) and
Eq. (\ref{pretercera}), respectively.
The empirical values of the power-law exponents 
will be obtained by means of maximum likelihood 
estimation 
applied in any case to both the rank-size relation $n(r)$ 
and the distribution of sizes $f(n)$.
Further, for illustrative and simplifying purposes,
we will compare the results of applying the ML-estimation formulas
for continuous power laws
(which is an approximate method for the discrete systems we are interested in)
with the results of ML estimation when the discreteness of the power-law
distributions is taken into account.
For continuous power laws we apply the method developed in Refs. 
\cite{Peters_Deluca,Corral_Deluca} (see also Ref. \cite{Corral_Gonzalez}).
The procedure for discrete power-law distributions is fully explained in our Appendix.
Motivation and an overview of both procedures follows.

\subsection{Maximum likelihood estimation and goodness-of-fit tests
\label{mle}}

The superiority of ML estimation in front of other methods of fitting
has already been pointed out by several authors.
In particular, for many years the most common used method, 
at least for fitting Zipf's law (and in complex-systems in general),
has been least-squares fitting. 
Important problems arise in this case when the empirical probability density 
or the empirical mass function
(for which the minimization with respect the fitting curve is performed)
are obtained using naive linear binning
 \cite{Goldstein,Pueyo,Bauke, White,Clauset}. 
Logarithmic binning of this function corrects some of these flaws
(when empty bins are not present),
as it does also the least-squares fitting of the cumulative
distribution (showing however other problems \cite{Hergarten_book,Burroughs}),
but still the least-squares method shows a considerable bias, 
high variance, and bin-size dependence, and yields distributions 
that are not normalized (as it does not use the fact that the 
curves to be fit are probability distributions).

In contrast, the ML estimator
(for distributions in the exponential family, 
where the power law belongs)
is the one with minimum variance among all
asymptotically unbiased estimators, 
a property which is called asymptotic efficiency \cite{pawitan2001,Bauke,White}.
Another advantage of this estimator is that it is invariant under reparameterizations
\cite{Casella,Corral_Deluca};
in other words, ML fits the distribution, not the parameters.
For all these reasons, ML estimation 
is employed in this article for the study of Zipf's law.

Given a continuous, non-truncated power-law distribution, 
$g(x) = (\tau-1) a ^{\tau-1}/x^\tau$,
defined for $x \ge a > 0$
with $g(x)$ the probability density of $x$
and $a$ the lower cut-off,
the ML estimation
of the exponent $\tau$ is straightforwardly obtained as
\begin{equation}
\hat \tau= 1 + \frac 1 {\ln G_a - \ln a },
\label{MLE1}
\end{equation}
where $G_a$ is the geometric mean of the values of $x$ in the sample 
fulfilling $x \ge a$, see Refs. \cite{Clauset,Corral_Deluca}.
For a discrete, non-truncated power-law distribution, $g(x) = 1/[\zeta(\tau,a) x^\tau]$,
defined for $x=a, a+1, \dots$ 
(with $g(x)$ the probability mass function
and $\zeta(\tau,a)$  the Hurwitz zeta function defined in the previous section),
the ML estimation of $\tau$
comes from the value that maximizes
the (per-datum) log-likelihood
\begin{equation}
\ell(\tau) = -\ln \zeta(\tau,a) - \tau \ln G_a.
\label{MLE2}
\end{equation}
In this case a closed solution for $\hat \tau$ is not possible, 
due to the presence of the function $\zeta(\tau,a)$ in the expression,
and one has to perform the maximization numerically
\cite{Clauset,Corral_Deluca_arxiv}.
It is clear that ML estimation for power-law distributions
is much simpler for continuous random variables, 
and for this reason it will be used here, together with the
more complicated discrete case 
(which is the natural procedure).
The random variable $x$ and the exponent $\tau$ will represent either 
the rank $r$ and the exponent $\alpha$ appearing in the version of
Zipf's law for types, Eqs. (\ref{laprimera}) or (\ref{laprimerabis}), 
or the size $n$ and the exponent $\gamma$ of Zipf's law for sizes, 
Eqs. (\ref{pretercera}) or (\ref{fn_sizes}).

In order to be more general, 
the power-law fitting is not performed in the full range $x\ge 1$
but for the upper tail of the distribution, 
whose starting point is given by the parameter $a$
(which could take the value $a=1$ as a particular case
and corresponds to $r_a$ and $n_a$ in each of the two representations).
This allows one to deal with empirical distributions that
are not pure power laws but only asymptotic power laws.
When the number of data is not infinite (i.e., always, in practice)
there will exist a value of $a$ for which an asymptotic power law
will be confused with a pure power-law;
the fact of fitting power-law tails takes advantage of this fact \cite{Voitalov_krioukov}.

As, a priori, $a$ is undetermined, 
one needs to do the fits for different values of $a$, and compare 
the goodness of each fit
(we sweep 20 values of $a$ per order of magnitude, equispaced in log-scale).
We take the smaller value of $a$ for which the fit is clearly non-rejectable
($p-$value greater than 0.2),
using the Kolmogorov-Smirnov goodness-of-fit test
for which the distribution of the Kolmogorov-Smirnov statistic
is calculated from 100 Monte-Carlo simulations \cite{Corral_Deluca,Corral_Deluca_arxiv}.
The simulations will allow us also to estimate the standard deviation
of the estimated value of the exponent.
This approach to power-law fitting and testing 
is inspired in Clauset et al.'s method \cite{Clauset}, 
but correcting some of its important shortcomings \cite{Corral_nuclear,Corral_Deluca}.


\subsection{Simulation of Zipf's law with rank as the random variable}

{Let us generate a synthetic sample of a Zipf's system, taking tokens 
from the types contained in a directory 
(e.g., word types in a special dictionary), 
which, in addition to the list of all possible
types also contains their global size or global frequency in the population.
We assume that
these global sizes $n$ come from a discrete power-law distribution, as
\begin{equation}
n(z) \propto \frac 1 {z^\alpha},
\label{EqX}
\end{equation}
for $z=1,2,\dots $,
where {$\alpha > 1$ and} $z$ denotes the values of a random variable associated to each type.
Notice that the random variable $z$ just represents an arbitrary labelling of the types
and it has no physical meaning
(except for its monotonic relation with $n(r)$).
This is the same model considered by Mandelbrot and mentioned above \cite{Mandelbrot61},
for which the number of possible types 
(number of possible values of $z$ in the population) is infinite.
If we draw a sample of $L_{tot}$ independent random numbers following the previous 
distribution we obtain a sequence of tokens, 
which constitutes our system of size $L_{tot}$.
By construction, this is a {\it Zipf's system}
obeying, in principle, Zipf's law for types.
The algorithm to simulate the discrete power-law distribution
is explained in the Appendix and
is a generalization of the one presented in Ref. \cite{Devroye}.

A plot showing the resulting of the process for a particular 
realization with $\alpha=1.2$ and $L_{tot}=10^6$ is displayed in Fig. \ref{appen1}.
The observed sizes 
$n$ of each type (or absolute frequencies, just counting tokens with
the same $z$) are plotted versus $z$;
in addition, a less naive estimation of the distribution of $z$
is obtained by adapting  
the logarithmic-binning procedure
explained in 
Refs. \cite{Christensen_Moloney,Deluca_npg}
to discrete distribution, see the Appendix.
(Note that for a power law without upper truncation
the value of $z$
can become colossal if the exponent $\alpha$ is close to one, 
as shown in Fig. \ref{appen1}, 
but this weird fact is not relevant in our argument.)

{However, in practice one has only access to the resulting sizes 
$n$ of the different types 
and not to the random variable $z$},
which we may consider then a hidden variable
{(i.e., a hidden rank with no physical representation that one can measure)}.
The substitution of $z$ by the rank $r$ is a useful trick,
performed ordering the resulting types (only those contained in the sequence)
by decreasing size $n$, and assigning rank values from 1 to $V_{tot}$
(remember that $V_{tot}$ is the total number of resulting types,
and equal to 133,146 in the realization of our example).
Figure \ref{appen1} shows how the rank is in good correspondence 
with the hidden variable $z$ for small values of $z$ 
(up to about 1000 for this concrete example),
but not for intermediate and large values, 
totally missing the long tail of $z$.

This failure of correspondence
is due to the unavoidable statistical fluctuations in finite samples, 
which make that although the probability of $z_1$ is higher than that of $z_2$
if $z_1 < z_2$, the value of $z_1$ does not appear necessarily more frequently 
than that of $z_2$ if $z_1$ is large enough; 
even more, $z_1$ may not appear at all in the sample. 
So, the rank is not a proper random variable.
{This can be seen more clearly from the definition of the random-variable
concept: one has to associate events to (natural) numbers for the whole sample space \cite{Kolmogorov1956aa}, but in the case of 
ranks the association is done a posteriori, after the random sample is drawn;
so, different samples lead to different assignations.}


\begin{figure}[!ht]
\begin{center}
\includegraphics[width=7.5cm]{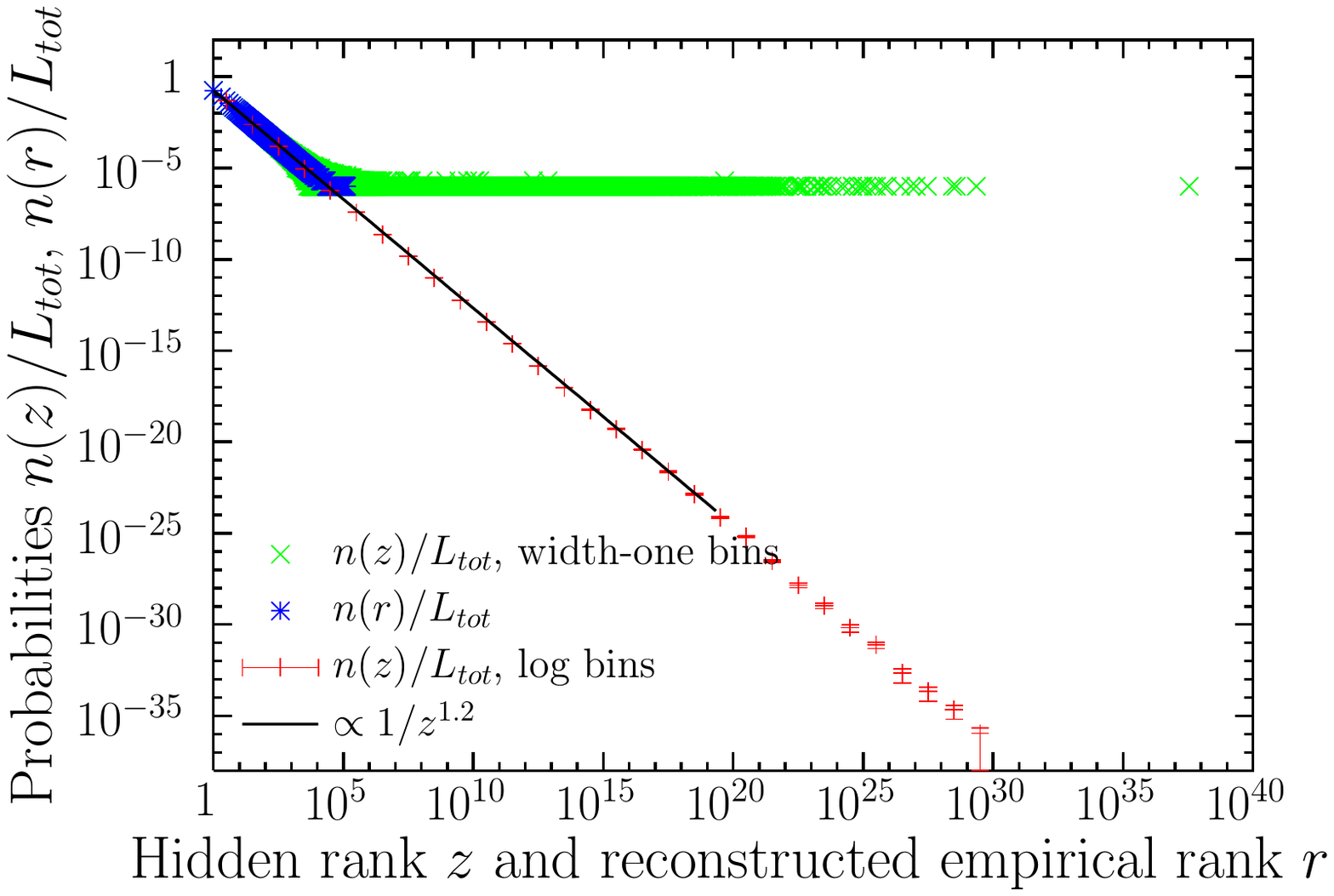}
\includegraphics[width=7.5cm]{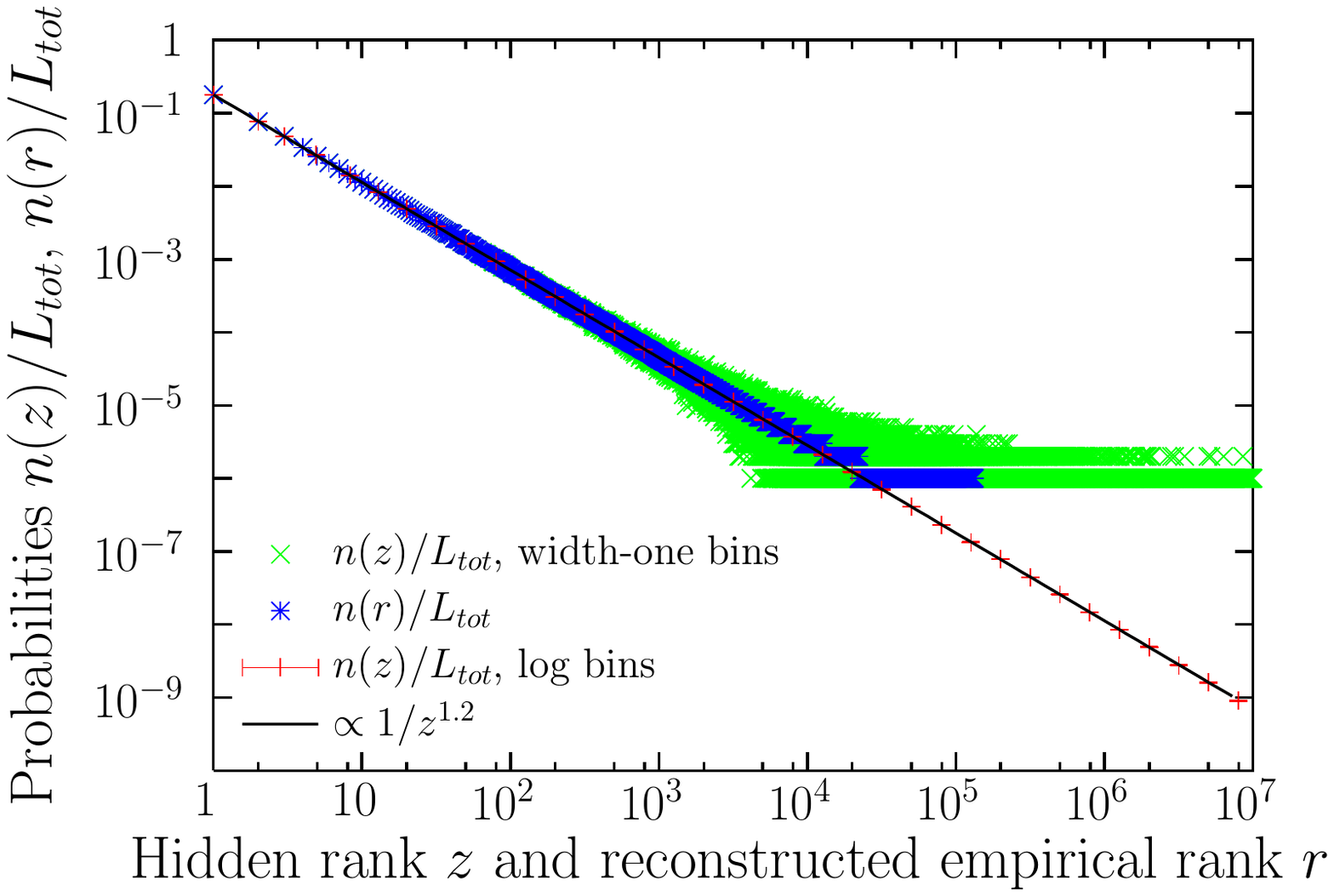}
\end{center}
\caption{
Estimation of the probability mass function $n(z)/L_{tot}$ 
of the hidden variable $z$ (hidden rank) associated to 
type size $n$ 
for a synthetic sequence with discrete power-law distribution of
$z$, Eq. (\ref{EqX}), with exponent $\alpha=1.2$ 
and length $L_{tot}=10^6$,
together with the corresponding empirical rank-size relation $n(r)/L_{tot}$.
Both size-one bins and logarithmic bins are shown for $n(z)/L_{tot}$.
The solid line is the original discrete power law from which
the tokens are drawn.
Notice how the only available quantity in practice, the rank-size
relation, deviates from a power law for large ranks
(corresponding to intermediate and large values of $z$).
This deviation (apart from the bias in the ML-estimated exponent)
is the responsible of the rejection of the power-law hypothesis.
The same distributions are shown at two different scales.
}
\label{appen1}
\end{figure}
 
\begin{figure}[!ht]
\begin{center}
\includegraphics[width=5in]{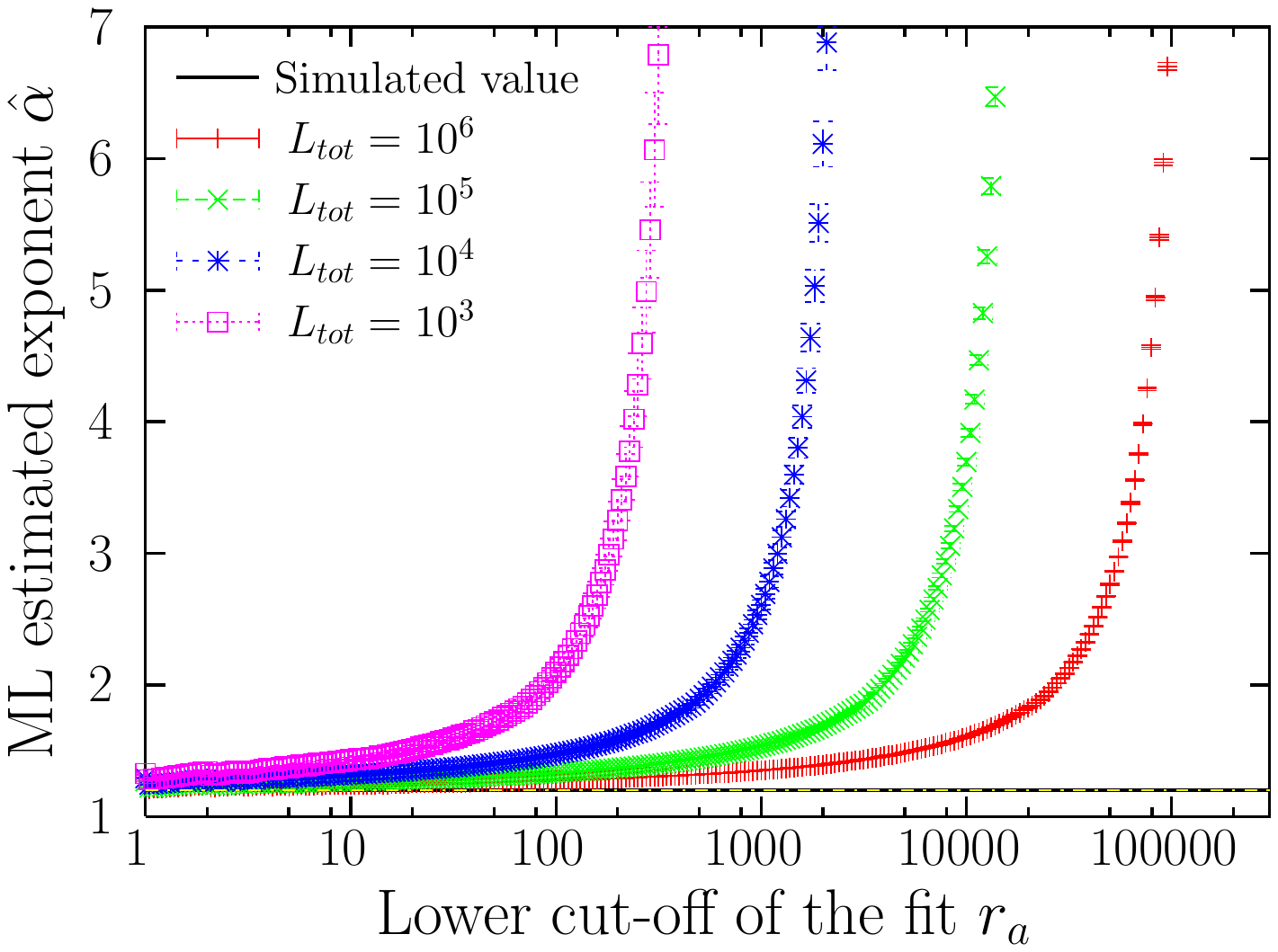}
\end{center}
\caption{
ML exponent from the rank-size representation, $\hat \alpha$, 
obtained  in the continuous approximation, 
as a function of the cut-off in rank, $r_a$, 
for several syntetic systems fulfilling Zipf's law for types, Eq.~(\ref{EqX}), 
with original value of $\alpha$ equal to 1.2
and different values of $L_{tot}$.
Observe the increasing positive bias from the true value with increasing $r_a$.
In no case the Kolmogorov-Smirnov test accepts 
the power-law hypothesis with the ML-estimated exponent.
Different values of the simulated $\alpha-$value lead to analogous results.
The fitting of a discrete power law yields essentially the same results
(not shown).
}
\label{appen1b}
\end{figure}

Let us apply 
the maximum-likelihood-estimation method,
together with the goodness-of-fit testing procedure referred to  
in the previous subsection, to the rank data obtained 
from the simulation.
To be precise, the quantities defined in the previous
subsection, $x$, $g(x)$, $\tau$, and $a$
correspond to $r$, $n(r)/L$, $\alpha$, and $r_a$, 
respectively. 
The outcome of the procedure 
leads to the rejection of the power-law distribution for $r$,
no matter the minimum value $r_a$ used to truncate the distribution from below,
and no matter also if a discrete or a continuous power law is fit;
that is, we do not find $p-$values larger than the 0.20 threshold 
for any value of $r_a$.
So, no power-law distribution can be fit to the rank-size
relation by ML estimation, although the relation 
is constructed from a true power law. 
For the particular example used for illustration, the case $r_a=1$
leads to 
$\hat \alpha=1.2214 \pm 0.0002$ for the discrete fit
and
$\hat \alpha=1.2516 \pm 0.0003$ for the continuous fit, 
close to the original value (with some positive bias)
but rejected for the reasons described below
(as a test of consistence, for the fit with the hidden rank $z$ and $r_a=1$ we get
$\hat \alpha=1.1994 \pm 0.0002$).

Figure \ref{appen1b} shows the positive bias of the ML estimation $\hat \alpha$
for different system sizes $L_{tot}$.
Indeed, a true power law would generate much higher values of the variable
(well beyond the 133,146 of maximum rank in our example), 
which would lead to larger geometric mean and to smaller exponents, through 
Eqs. (\ref{MLE1}) or (\ref{MLE2}).
So, the fact that $r<z$  for large $z$
leads to an underestimation of the geometric mean of the variable
and to an overestimation of the exponent.
The bias is specially pronounced for large $r_a$,
corresponding to a decreasing number of data in the power law.
Nevertheless, the rejection of the power law is achieved through the Kolmogorov-Smirnov test, due to the lack 
of resemblance of the rank data with true power-law distributed data.
Visual inspection of the rightmost part of Figure \ref{appen1} seems to indicate that
a ``flat'' power law, i.e., one with an exponent equal to zero, 
could fit the largest ranks (those with $n=1$); nevertheless, 
such distribution is not normalizable when defined over an infinite support
and cannot arise therefore 
from the ML formalism
(except if one introduces an upper truncation in the power law).
On the other side, 
one could envisage a goodness-of-fit test in which empirical rank-size data
are compared with simulated rank-size data.
This is not contemplated in the standard algorithms provided in Refs. 
\cite{Clauset, Corral_Deluca} and it will be the subject of future research.

On the contrary, application of ML estimation to the distribution of type 
sizes $f(n)$ (counting how many types have a given size) leads to 
the acceptance (that is, no rejection) of the power-law hypothesis
for precise values of the lower cut-off $n_a$.
The correspondence with the quantities of subsection \ref{mle}
is $x= n$, $g(x)=f(n)$, $\tau=\gamma$, and $a=n_a$. 
In the concrete example mentioned above with $\alpha=1.2$ we get,
for discrete ML estimation,
an estimated power-law exponent $\hat \gamma=1.86 \pm 0.01$
starting to hold for $n \ge 7$, with a $p-$value 0.85.
In the simpler continuous approximation the results
are $\hat \gamma=1.84 \pm 0.02$
for $n \ge 32$, with a $p-$value 0.29.
The corresponding values of $\alpha$ are,
using Eq. (\ref{gammabetaalpha}),
$\hat \alpha=1.16$
and
$\hat \alpha=1.19$, respectively, so, there is some bias 
(the true exponent of the asymptotic power law
is $\gamma=1+1/1.2=1.833$, after Eq. (\ref{gammabetaalpha})).
Both fits are shown in Fig. \ref{fig_tokens}(a) in terms of $f(n)$
and $S(n)$.
The ``translation'' of the fits into the rank-size format
appears in Fig. \ref{fig_tokens}(b).
Table \ref{tableone} summarizes more results from simulations of this kind.
A total of 20 systems are simulated from Zipf's law for tokens
for each value of $\alpha$,  being these $\alpha=1.2, 1.3$, and $1.4$.
In no case there is a contradiction with the conclusions obtained from
the example shown in Fig. \ref{fig_tokens}.

\begin{table}[!ht]
\begin{center}
\caption{\label{tableone} 
Maximum likelihood fitting of $f(n)$ 
for synthetic systems fulfilling (by construction) the Zipf's law for types, Eq. (\ref{EqX}).
For each value of $\alpha$ we generate 20 systems with $L_{tot}=10^6$ tokens each, 
and both discrete ML estimation (upper row for each $\alpha$) 
and continuous ML estimation (lower row) are performed.
The averages (represented by the bar) of the estimated exponent $\hat \gamma$, 
of the cut-off  $n_a$, and of the $p-$value, are over the 20 samples, 
and the error is one standard deviation of the variable (not of its mean, 
which is a factor $\sqrt{20}$ smaller).
Also, the average and the 
estimation of the standard deviation of the resulting
number of types $V_{tot}$ are included.
}
\smallskip
\begin{tabular}{|rr|rrrr|l|}
\hline
&&&&&&\\ 
$\alpha$ & $\gamma$ & $\bar V_{tot}$  & $\bar {\hat \gamma} $ & $\bar n_a $ & $\bar p$ & ML estimation \\
\hline
1.20 & 1.833 & 132,934$\pm$258& 1.861$\pm$0.010    &7.1$\pm$3.0 & 0.51$\pm$0.28 & discrete \\
 &&                                           & 1.842$\pm$0.007  & 32.5$\pm$3.2 & 0.47$\pm$0.19 & continuous\\
\hline
1.30 & 1.769 &  56,771$\pm$168 & 1.794$\pm$0.009  &   6.0$\pm$1.6 & 0.58$\pm$0.25 & discrete\\
 &&                                          & 1.774$\pm$0.006  &  25.3$\pm$4.1 & 
0.38$\pm$0.14  & continuous\\
\hline
1.40 & 1.714&  27,098$\pm$ 88  & 1.739$\pm$0.007  &    5.0$\pm$1.2& 0.57$\pm$0.25& discrete\\
&&                                           & 1.722$\pm$0.008  &    22.8$\pm$3.4& 0.42$\pm$0.17  & continuous\\
\hline
\end{tabular}
\end{center}
\end{table}

These results lead to the remarkable situation that although the underlying pure
Zipf's law may be valid for types, we find the distribution of sizes $f(n)$
more reliable than the statistics of types $n(r)$ in order to test the power-law hypothesis
when the best method of parameter estimation (ML, as explained in the previous
subsection) is used \cite{Corral_Deluca}.
In fact, the approximate continuous procedure seems to yield better results than
the exact discrete case. 
The reason is that the true empirical distribution of sizes is only a power-law
asymptotically, and presents a strong excess of probability for the smallest values of $n$ 
(in comparison with a pure power law). 
As the continuous ML method 
works worse for discrete data, it rejects the power-law hypothesis
for small values of $n_a$, avoiding the deviations from the power law and yielding an exponent closer to the original one.
In contrast, the discrete ML method is able to accept the hypothesis
for smaller $n_a$'s, at the price that there is a certain bias in the value of the exponent.
In practice, both fits look very satisfactory, as shown in Fig. \ref{fig_tokens},
although, due to its smaller bias one may prefer the continuous case. 
Figure \ref{fig3bis} shows the biased result for the exponent $\hat \gamma$
for small values of the lower cut-off $n_a$.
Note that it is not the ML method that leads to biased results, 
but that the distribution deviates from a power law when small values of $n$
are taken into account, as seen in Fig. \ref{fig_tokens}(a).

\begin{figure}[!ht]
\begin{center}
(a)
\includegraphics[width=7.5cm]{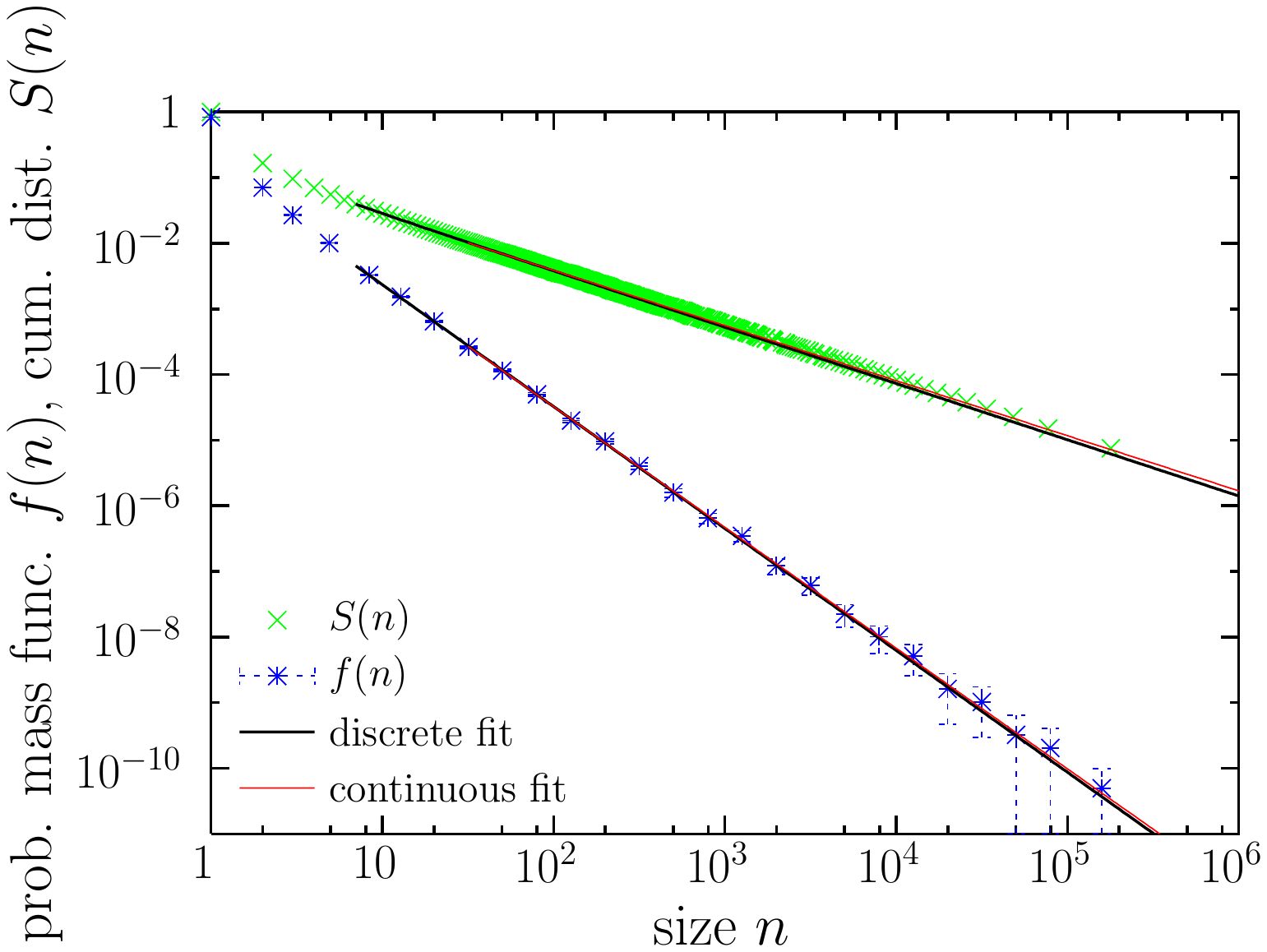}
(b)
\includegraphics[width=7.5cm]{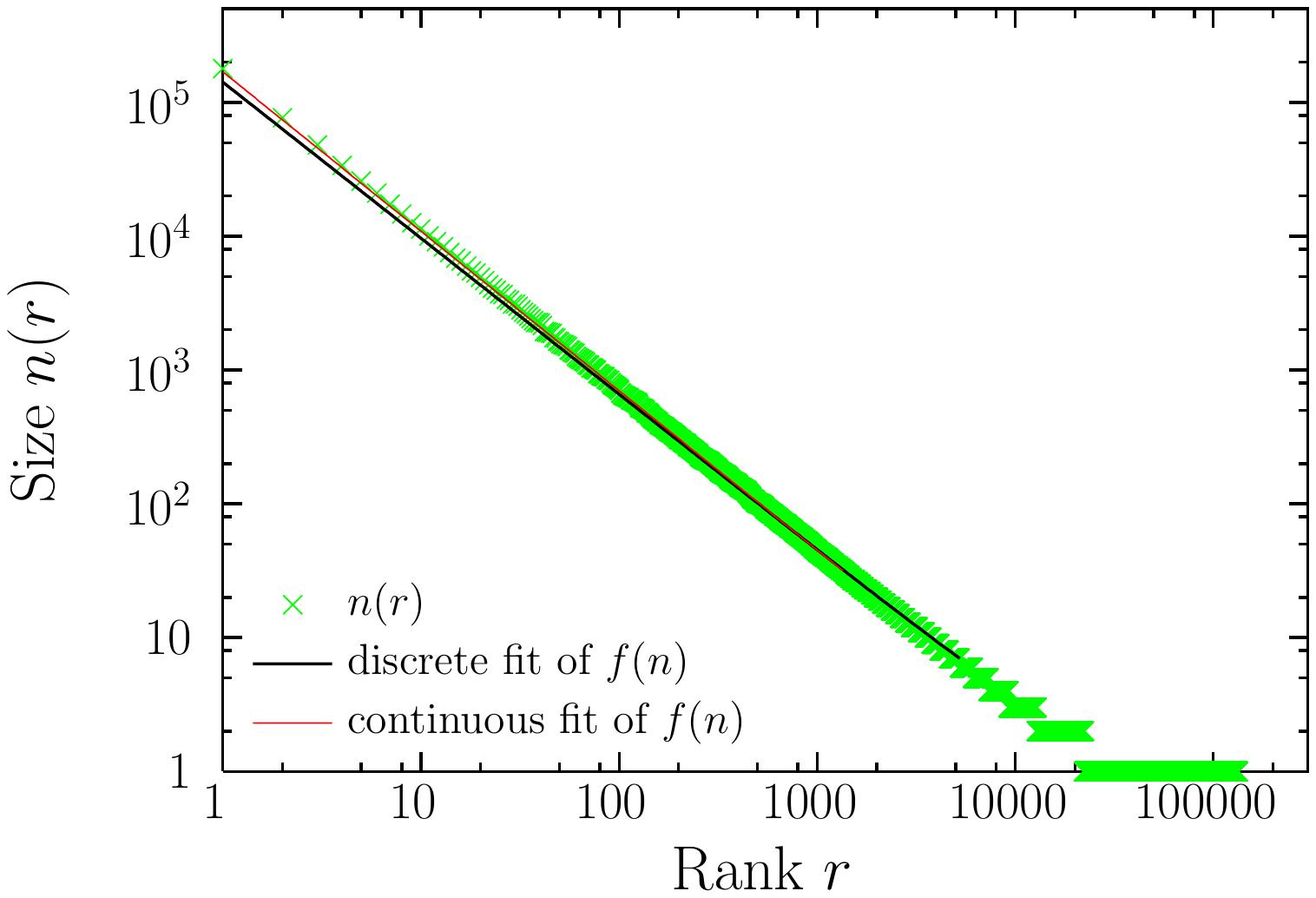}
\end{center}
\caption{
(a) Estimated probability mass function $f(n)$ 
and survivor function $S(n)$
for simulated Zipf's law for types, Eq. (\ref{EqX}), 
taking $\alpha=1.2$ and $L_{tot}=10^6$
(same data as in Fig. \ref{appen1}),
together with the corresponding power-law ML fit for the random variable $n$, 
Eq. (\ref{fn_sizes}),
in the discrete case (black line)
and also in the continuous approximation (red line).
The fit in terms of $S(n)$, given by
Eq. (\ref{Sn_sizes}), is also shown. 
We remark that ML estimation does not rely on the
estimations of $f(n)$ of $S(n)$,
these are shown here as a visual verification of the goodness
of the fits.
Expression (\ref{Mandelbrot_Simon}) 
provides a good fit of $f(n)$ for all $n$ (not shown).
(b) Translation of the previous ML fit of the $n$ variable
into the rank-size representation, given by
the inverted Hurwitz zeta function
of Eq. (\ref{Hurwitz_inversion}).
The corresponding values of $r_b$
turn out to be $r_b\simeq 5250$ (discrete fit)
and $r_b\simeq 1350$ (continuous fit).
}
\label{fig_tokens}
\end{figure}

\begin{figure}[!ht]
\begin{center}
\includegraphics[width=12cm]{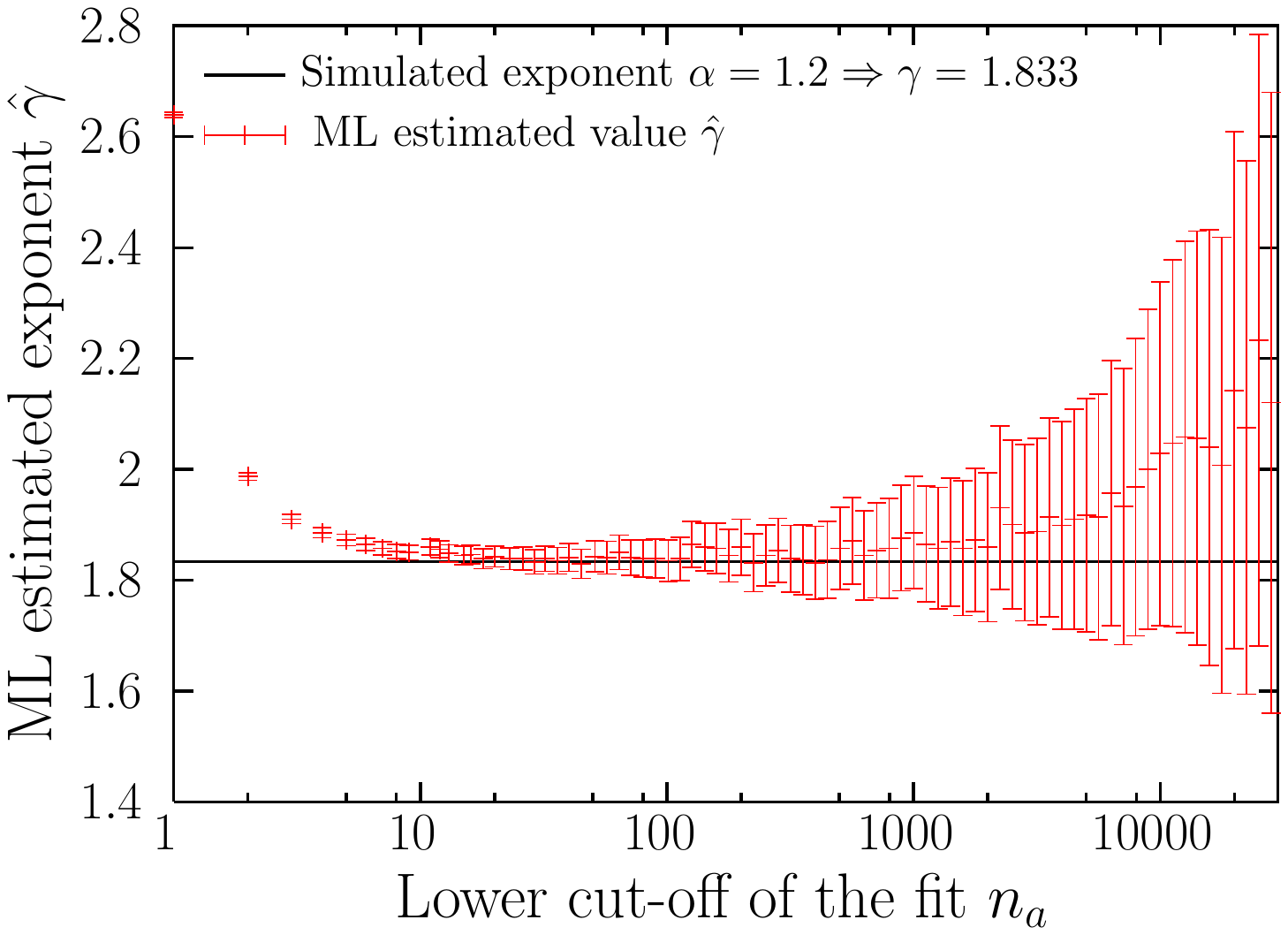}
\end{center}
\caption{
ML exponent from the distribution-of-sizes representation, $\hat \gamma$,
as a function of the cut-off in size, $n_a$, 
for a synthetic system fulfilling Zipf's law for types, Eq.~(\ref{EqX}), 
using the (exact) discrete fitting procedure.
The original value of $\alpha$ is equal to 1.2 
(represented as a straight line for $\gamma=1.833$)
and $L_{tot}=10^6$
(same data as in Fig. \ref{fig_tokens}).
The continuous fitting leads to similar results, although
in this case the fit is not accepted for larger values of $n_a$.
}
\label{fig3bis}
\end{figure}

\subsection{Simulation of Zipf's law with size as the random variable}

Now we generate a synthetic Zipf's law not for types (as in the previous subsection) 
but for sizes, 
i.e., the discrete power-law distribution for $f(n)$, Eq. (\ref{fn_sizes}), 
holds exactly in the population.
Thus (in order to compare with that subsection), 
we generate $V_{tot}=133,000$ independent random numbers from 
a discrete power law, $f(n)=1/[\zeta(\gamma) n^{\gamma}]$, 
defined for $n=1, 2, \dots$ (i.e., $n_a=1$), 
with 
$\zeta(\gamma)=\zeta(\gamma,1)$
the Riemann zeta function
and
exponent $\gamma=1.833$ (corresponding to $\alpha=1.2$), 
using the algorithm explained in the Appendix.
Each of these random numbers represents the size $n$ of a type, 
and so, $f(n)$ is obtained directly from the statistics of $n$.
To plot the rank-size relation we order the list of types by decreasing
value of $n$, and assign ranks $r=1,2,\dots V_{tot}$.

In order to 
apply ML estimation to the rank variable 
we need to generate a synthetic system of tokens
(the key step is the calculation of the empirical value of the 
geometric mean of the random variable $r$
Eqs. (\ref{MLE1}) and (\ref{MLE2})), 
this is done by creating $n(r)$ copies of each of the $V_{tot}$ types,
each labeled by its corresponding rank $r$.
This list of $L_{tot}$ values 
is the data entering as the input of the ML routine 
in the rank-size approach.
This procedure leads to the same results
as in the previous subsection, that is, the power-law hypothesis for the rank-size
relation is rejected,
no matter how large the value of the lower cut-off $r_a$ is.

In contrast, ML for the random variable $n$ leads to the acceptance of 
the existence of a power-law distribution.
This is obvious, as $n$ comes indeed from a power-law distribution, 
for which, at variance with the Zipf's law for types, 
the value of its random variable is not hidden. 
The results for the selected example (equivalent to Fig. \ref{appen1}) 
are $\hat \gamma=1.835 \pm 0.004$
for $n \ge 2$, with a $p-$value 0.31 for ML estimation
of a discrete power law and
$\hat \gamma=1.838 \pm 0.015$
for $n \ge 56$, with a $p-$value 0.23
for the continuous-version approximation.
Figure \ref{fig_types}(a) shows the direct outcome
of the fit, both in terms of $f(n)$ and $S(n)$, whereas 
Fig. \ref{fig_types}(b) shows how the fit of the distribution of $n$
translates into the rank-size representation.

Note that the resulting system size, $L_{tot}=\sum_{r=1}^{V_{tot}} n(r)$,
turns out to be, in the particular realization chosen as an example, 
$L_{tot}=3,417,385$.
The large difference with the value $L_{tot}=10^6$ used in the previous subsection 
for about the same $V_{tot}$ is due to
the distinct balance between types with $n=1$ and the rest.
In the simulation using Zipf's law for types
there was an excess of very small $n-$values;
so, for the same $V_{tot}$ the size of the system $L_{tot}$ becomes smaller there.
Figure \ref{comparison} shows together, in order to ease the comparison, 
the $f(n)$ resulting from simulating Zipf's law for types (previous subsection)
with Zipf's law for sizes (this subsection).

Another source of variation in the value of the resulting system size $L_{tot}$
is that this arises as a sum of independent power-law distributed frequencies $n$.
As the exponent of the power law $\gamma$ is smaller than 2, 
the law of large numbers does not apply and the sum is not scaling linearly
with the number of terms (types) $V_{tot}$.
Instead, the sum is broadly distributed, as the generalized central limit theorem teaches us
\cite{Bouchaud_georges,Corral_csf,Corral_Font_Clos_PRE17}. 
Table \ref{tabletwo} provides the results obtained from other examples simulating 
Zipf's law for sizes [Eq. (\ref{fn_sizes})]; 
these results are in total agreement with the example
chosen for illustration and show the wild dispersion in the resulting values of $L_{tot}$.
In this case it is clear that the discrete fit is preferred, 
as it leads to a much smaller value of $n_a$,
which yields more data in the power-law regime and
a smaller uncertainty in the exponent.

\begin{table}[!ht]
\begin{center}
\caption{\label{tabletwo} 
Equivalent to Table \ref{tableone}, for systems fulfilling Zipf's law for sizes.
For each value of $\gamma$ we generate 20 systems with $133,000$ types each.
In this case it is clear that discrete ML estimation provides better results
than the continuous approximation.
Notice the enormous variation on $L_{tot}$,
described by its minimum and maximum values.
}
\smallskip
\begin{tabular}{|rr|rrrrr|l|}
\hline
$\gamma$ & $\alpha$ & min $L_{tot}$ & max $L_{tot}$ & $\bar {\hat \gamma} $ & $\bar n_a $ & $\bar p$ & ML estimation \\
\hline
1.833 &  1.20 & 2.9 $\cdot 10^6$ & 2.4 $\cdot 10^8$& 1.833$\pm$0.003 &   1.2$\pm$ 0.5 & 0.61$\pm$0.24 & discrete\\
 &&   &                                                  & 1.834$\pm$0.018 & 56.2$\pm$23.9 & 0.32$\pm$0.13 & continuous\\
1.769 &  1.30 & 7.2 $\cdot 10^6$ & 1.2 $\cdot 10^9$ & 1.769$\pm$0.003 &  1.4$\pm$ 0.9 & 0.61$\pm$0.23 & discrete\\
 &&  &                                                   & 1.773$\pm$0.012 & 55.4$\pm$17.4& 0.38$\pm$0.14 & continuous\\
1.714 &  1.40 & 1.6 $\cdot 10^7$ & 5.9 $\cdot 10^9$ & 1.713$\pm$0.002 &  1.3$\pm$ 0.4 & 0.66$\pm$0.26 & discrete\\
&&    &                                                   & 1.716$\pm$0.012 & 63.0$\pm$23.8    & 0.35$\pm$0.17 & continuous\\
\hline
\end{tabular}
\end{center}
\end{table}


\begin{figure}[!ht]
\begin{center}
(a)
\includegraphics[width=7.5cm]{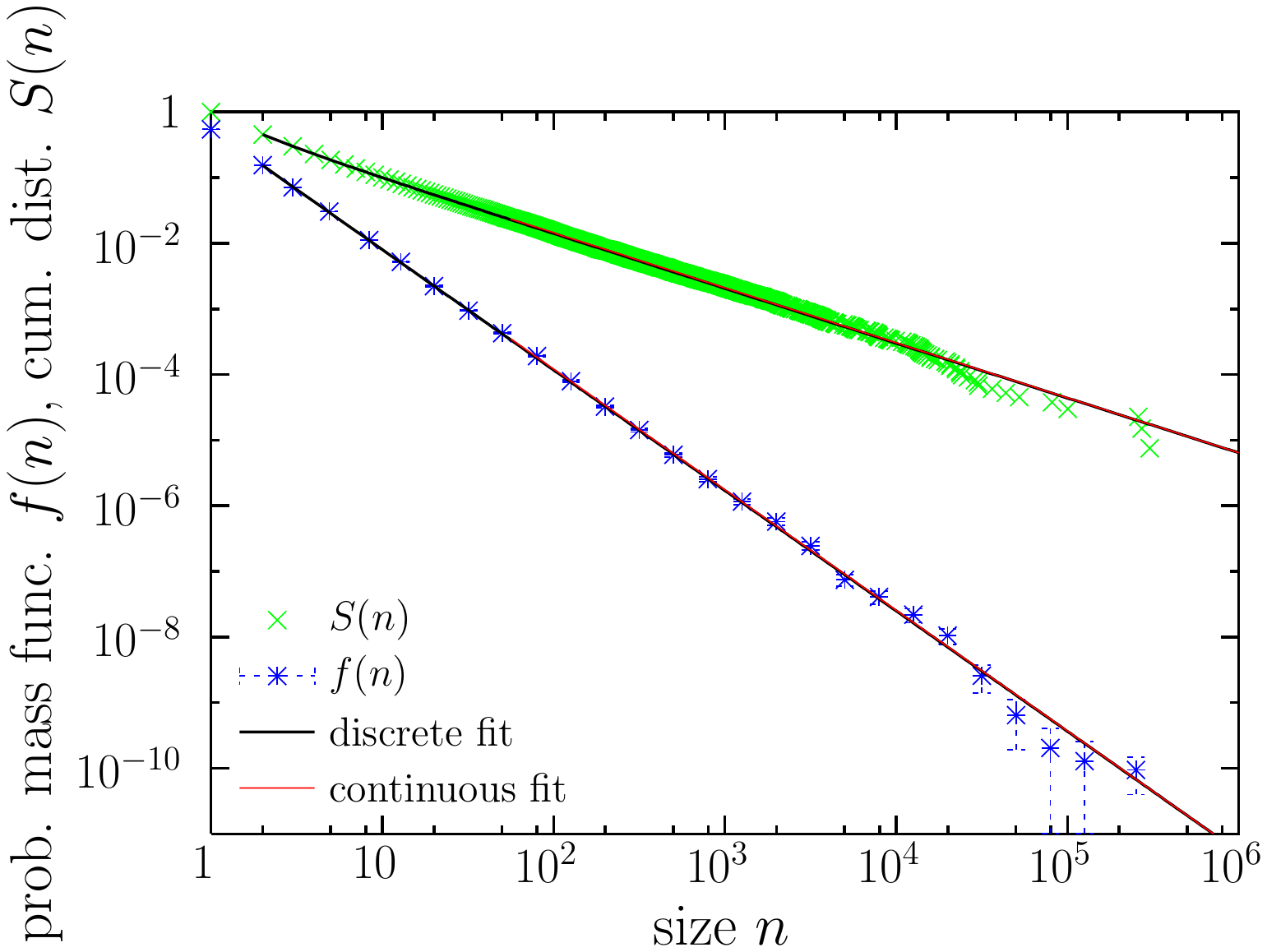}
(b)
\includegraphics[width=7.5cm]{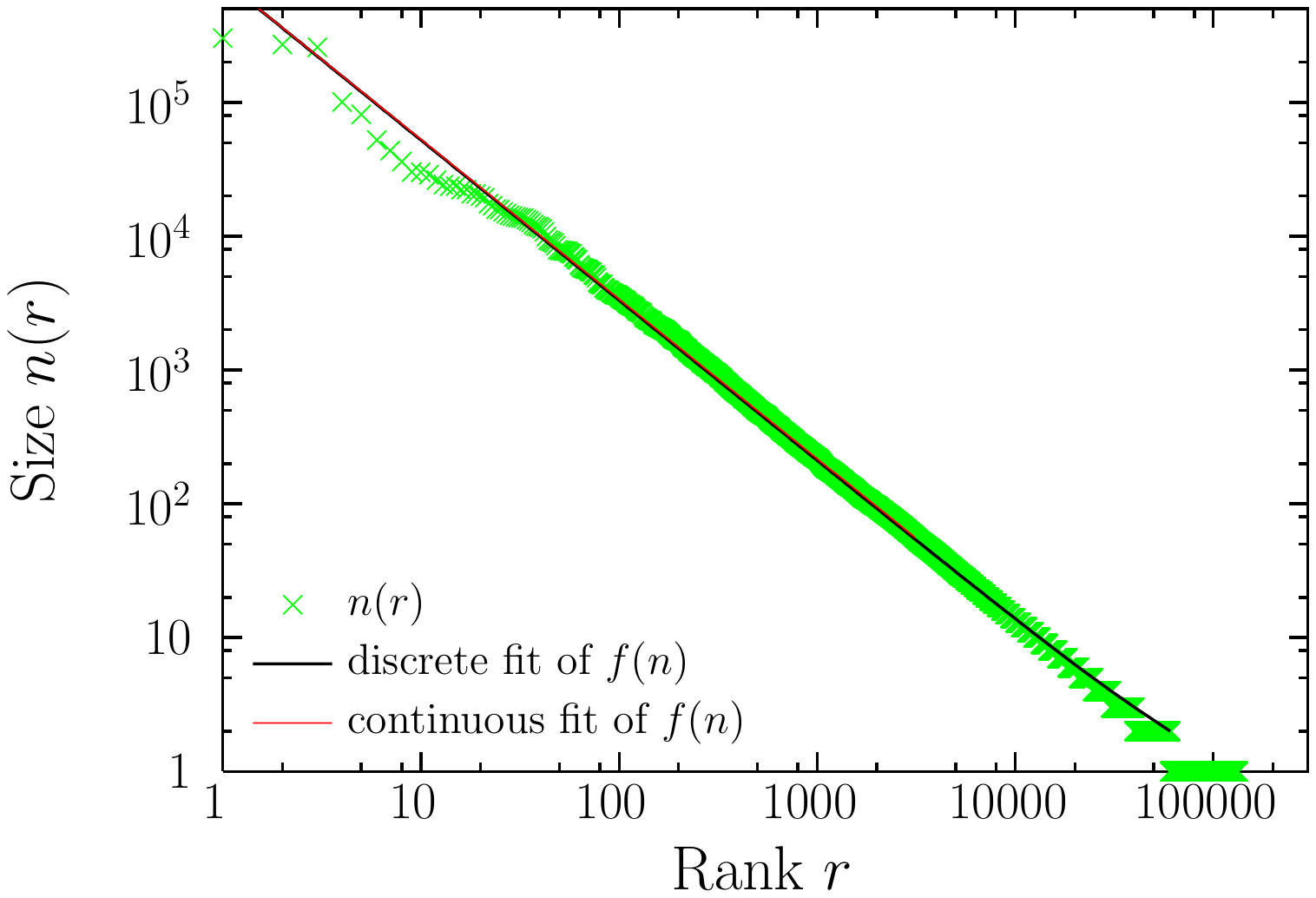}
\end{center}
\caption{
Same as 
Fig. \ref{fig_tokens} replacing the simulation of the Zipf's law for types
by the simulation of the Zipf's law for sizes, Eq. (\ref{fn_sizes}),
with $\gamma=1+1/1.2\simeq 1.83$, $n_a=1$, and $V_{tot}=133,000$.
(a) Empirical cumulative distribution and probability mass function of sizes, 
together with discrete and continuous fits.
(b) Empirical rank-size representation and fits. 
}
\label{fig_types}
\end{figure}

\begin{figure}[!ht]
\begin{center}
\includegraphics[width=13.5cm]{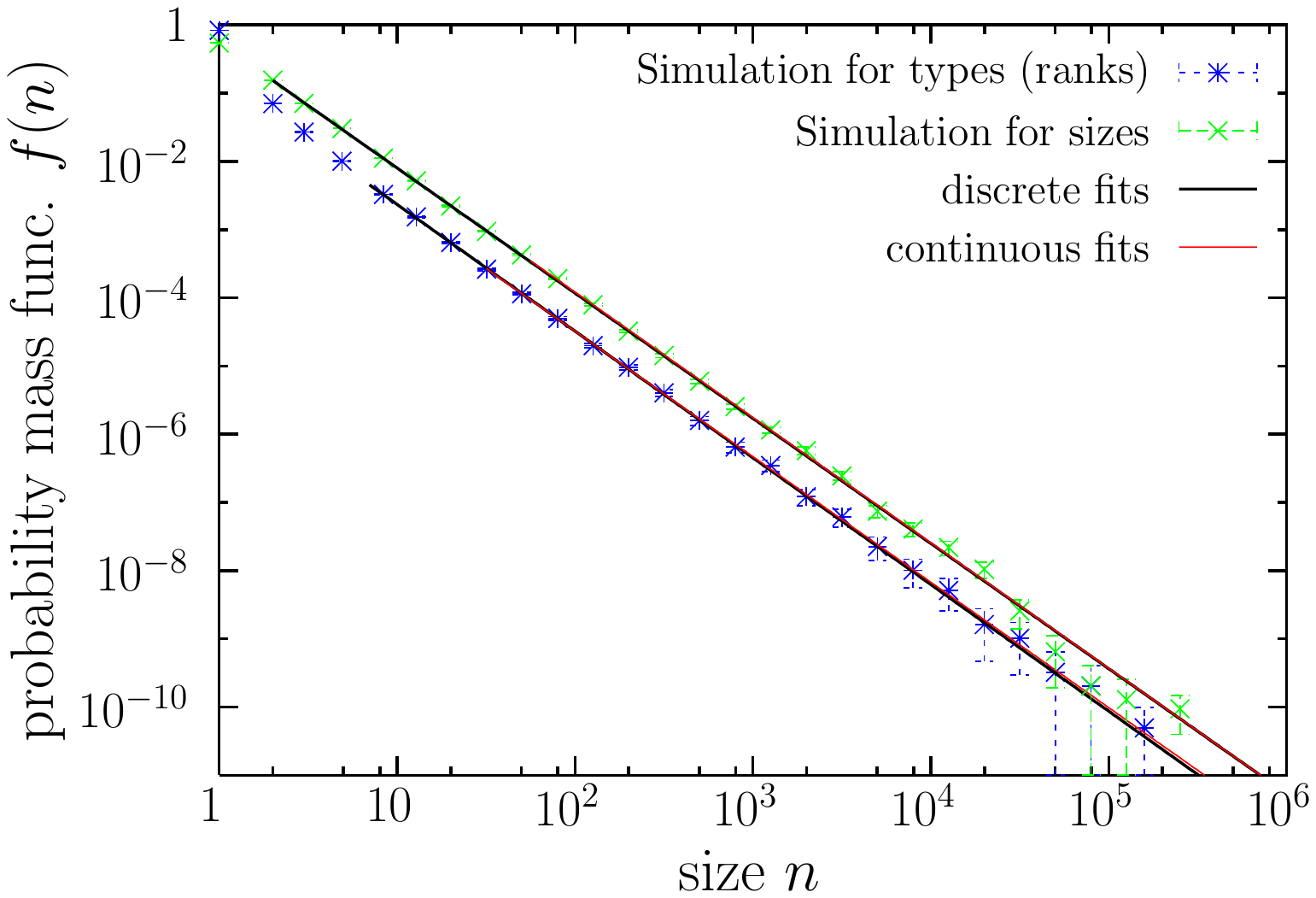}\\
\end{center}
\caption{
Direct comparison of the estimated probability mass functions $f(n)$
simulating both versions of Zipf's law, 
Eq. (\ref{EqX}) [Fig. \ref{fig_tokens}(a)] and 
Eq. (\ref{fn_sizes}) with $n_a=1$ [Fig. \ref{fig_types}(b)].
Notice the bending of $f(n)$ for small $n$ when Zipf's law is simulated
for types.
}
\label{comparison}
\end{figure}

\section{Discussion}


We have shown, empirically, the clear advantages of looking for the fulfillment
of Zipf's law by using maximum likelihood estimation applied to the distribution of sizes, 
instead of to the rank-size relation.
Table \ref{ultimatablalojuro} summarizes our results.
Other arguments in favor of the distribution of sizes can be the following:

\begin{itemize}

\item
{\em Parsimony (number of parameters).}\\
According to standard model selection, the evaluation of the goodness of a model 
(e.g., a power law) must be based not only on the quality of the fit of the model 
but also on a penalty for the number of parameters employed \cite{Burnham2002a}. 
The fact that 
$\alpha = 1/(\gamma -1)$ 
{(recall Eq. \ref{gammabetaalpha})}
means that if $\gamma \geq 2$ then $\alpha \leq 1$. 
These values of the exponent have been claimed for human language 
under certain conditions 
{\cite{Ferrer2004a}}.
The problem is that a power-law distribution 
needs $\alpha > 1$ {for normalization}
{(otherwise, the summation of the probabilities does not converge).}
Therefore, when sizes are taken {for statistical modeling, just two parameters
suffice for a power-law description: 
$\gamma$ and a minimum cut-off $n_a$; in contrast, 
the corresponding power-law for ranks needs three parameters: 
$\alpha$, a minimum rank $r_a$, and a maximum rank $r_b$.}
In fact, 
this maximum rank is unavoidable in practice, even for $\alpha > 1$,
due to the distortions introduced in the rank-size relation by the artifactual nature of the rank
(as we have seen, because of the plateaus corresponding to the smallest frequencies, which lead to the 
rejection of the non-truncated power-law hypothesis for ranks).
We have not considered truncated power laws in this article 
{\cite{Baixeries2012a}}, 
but it is clear from our results that their use for the rank-size relation must be accompanied by
model selection criteria such as BIC, AIC, etc. 
{\cite{Burnham2002a}}, if a comparison with the representation in terms
of the distribution of sizes is going to be performed. 

\item
{\em Bias for a negative correlation.}\\
The definition of rank forces the size (i.e., the number of tokens) 
of a rank to not increase as rank increases. 
This {implies} a negative correlation between a rank and its size. 
In contrast, the number of types with a certain size is 
free with regard to size, in principle: it can increase, decrease or remain constant as 
size increases. 
This {difference} is vital for model testing as 
the null hypothesis of a power law might be harder to reject 
{in terms of the rank-size relation 
because of this correlation.
The situation is analogous to the problems arising when fitting
a probability distribution from its cumulative distribution function
\cite{Hergarten_book}.}
In our case, although we get $p-$values equal to zero
when a non-truncated power law is fit to the rank-size relation,
the opposite effect, leading to inflated $p-$values, may arise
if we truncate the rank-size power law at a maximum value $r_b$.
Indeed, we have some preliminary results indicating that
between $r=100$ and $1000$ the $p-$values of a rank-size relation
generated from a discrete power law with exponent $\alpha=1.2$ 
are not uniformly distributed
between $0$ and $1$ but biased to the high values.

\begin{table}[!ht]
\begin{center}
\caption{\label{ultimatablalojuro} 
Summary of results found in this article.
PL stands for the null hypothesis of a power-law distribution.
}
\smallskip
\begin{tabular}{|l|l|l|}
\hline
Simulating &ML applied to rank $r$ & ML applied to size $n$\\
\hline
Zipf's law for types, Eq. (\ref{laprimerabis})&PL rejected& PL not rejected \\
&&(slight positive bias of $\hat \gamma$\\
&&\phantom{(}for the discrete fit)\\
\hline
Zipf's law for sizes,  Eq. (\ref{fn_sizes})     &PL rejected& PL not rejected\\
&&(too large value of $n_a$\\
&&\phantom{(}in the continuous fit)\\
\hline
\end{tabular}
\end{center}
\end{table}

%
%
%
\end{itemize}

An opposite argument (against the distribution of sizes and in favor of the rank-size relation)
can be that the distribution-of-sizes approach entails a substantial reduction in the number of data
when the texts are large. 
Indeed, Heaps' law \cite{Baayen,FontClos_Corral}
approximately relates vocabulary (empirical number of types)
with text length (empirical number of tokens) as 
$V_{tot}\propto L_{tot}^{1/\alpha}$, if $\alpha >1$, 
which implies $V_{tot}\ll L_{tot}$ if $L_{tot}$ is large 
(we have clearly seen this in the example chosen for several figures, 
with $L_{tot}=10^6$ and $V_{tot}\simeq 133,000$).
This can make the power-law hypothesis more difficult to reject in the distribution-of-sizes representation. Nevertheless, the advantages shown in this article 
for the distribution-of-sizes representation overcome this little disadvantage.

In summary, we have presented wide evidence that the description of Zipf's power law
is a different matter in terms of the rank-size relation 
and in terms of the distribution of sizes.
In other words, both descriptions lead to different
distributions of tokens into types.
Whatever version of Zipf's law might hold in real systems, 
or even, if neither of the two versions is expected to hold, 
the application of maximum likelihood estimation has to be done 
taking the size as the random variable.
We recommend working therefore
in the distribution-of-sizes representation.

\section{Acknowledgements}

We thank 
P. Puig for discussions about statistical testing, 
{G. Altmann and G. Boleda for his advice from a quantitative linguistics perspective,}
L. Devroye for facilitating a copy of his valuable book for free,
and A. Deluca and F. Font-Clos for 
feedback on the maximization of $g(m)/q(m)$, 
which should complement Devroye's book.
Support from projects
FIS2012-31324, FIS2015-71851-P, 
Mar\'{\i}a de Maeztu Program 
MDM-2014-0445,
from Spanish MINECO,
and the Collaborative Mathematics Project from La Caixa Foundation
(I.S.) is acknowledged.


\section{Appendix}

\subsection*{Fitting and testing discrete power laws}

We now explain the procedure of finding accurate values of the parameters
that describe the discrete power-law distribution
{Our method is based in the one by
Clauset {\it et al.} \cite{Clauset},
but introducing important modifications that yield
a better performance \cite{Peters_Deluca,Corral_nuclear}.
As the continuous case is treated in those references
we explain here the peculiarities of the discrete fitting.
In the exposition we use a generic representation 
that is valid both for the rank-size representation and
for the distribution of sizes.
Table \ref{equivalence} provides an equivalence between the notation 
used in each representation.

\begin{table}[h]
\begin{center}
\caption{\label{equivalence} 
Correspondence of notation between the generic representation used 
in this section (and in Subsec. \ref{mle}),
the rank-size representation, 
and the representation in terms of the distribution of sizes.
Being strict, we should have defined a function $h(r)$ as $h(r)=n(r)/L$, 
but we have preferred to avoid a growth in the notation.
}
\smallskip
\begin{tabular}{|l| cccccc|}
\hline
&&Mass &Cumul. &&Lower & Number\\
&Variable & func. &distrib.& Exponent &cut-off& of data\\
\hline
Generic representation & $x$ & $g(x)$ & $G(x)$ &$\tau$ & $a$ & $N_a$ \\   
Rank-size representation & $r$ & $n(r)/L$ & -- &$\alpha$ & $r_a$ & $L$ \\
Distribution of sizes & $n$ & $f(n)$ & $S(n)$ &$\gamma$ & $n_a$ & $V$ \\
\hline
\end{tabular}
\par
\end{center}
\end{table}

}

Let us consider a discrete power-law distribution, 
defined for $x\ge a$,
with $a$ a natural number ($a\ge 1$).
The corresponding probability mass function is
$$
g(x) = \frac 1 {\zeta(\tau,a) x^{\tau}}, \, \mbox{ for } \, x=a,a+1,a+2,\dots
$$
(and zero otherwise),
where normalization is ensured by the Hurwitz zeta function, defined as
$$
\zeta(\tau,a) = \sum_{k=0}^\infty \frac 1 {(a+k)^\tau};
$$
for $a=1$, $\zeta$ becomes the standard Riemann zeta function,
and the distribution is called the Riemann zeta distribution
(or rather confusingly, the discrete Pareto distribution \cite{Johnson_univariate}).
The corresponding (complementary) cumulative distribution function is
$$
G(x)=\frac{\zeta(\tau,x)}{\zeta(\tau,a)},
$$
for $x\ge a$, giving, by definition, $\mbox{Prob}[\mbox{variable} \ge x]$.
Our approach fits the value of $\tau$ corresponding to different values of $a$
and selects the one that yields {the largest power-law range}
provided that the quality of the fit is acceptable, as explained next.

\subsubsection*{ML estimation and computation of the Hurwitz zeta function}

The first step (step 1) then is the fitting of $\tau$. The method we use
is maximum likelihood estimation.
Considering $a$ as a fixed parameter,
the (per-datum) log-likelihood function $\ell$ for a discrete power-law distribution is
defined as the logarithm of the likelihood function,
divided by the total number of data 
$N_a$
in the power-law range (i.e., those for which $x\ge a$);
this is,
$$
\ell(\tau) = \frac 1 {N_a}  \ln \prod_{i=1}^{N_a} g(x_i)=
 \frac 1 {N_a} \sum_{i=1}^{N_a} \ln g(x_i),
$$
with $x_i$ the recorded values of $x$,
numbered from $i=1$ to $N_a$.
Values below $a$ must be disregarded.
This yields
$$
\ell(\tau) = -\ln \zeta(\tau,a) - \tau \ln G_a,
$$
where $G_a$ is the geometric mean of the data in the range, 
that is, 
$$\ln G_a =\frac 1 {N_a} \sum_{i=1}^{N_a} \ln x_i$$ for $x_i\ge a$.
As $a$ and the data are constants, $\ell$ is only a function of $\tau$,
and the maximum of this function yields the estimation of $\tau$,
which we call $\tau_{emp}$, that is,
$$
\tau_{emp}= \mbox{arg}\max_{\forall \tau} \ell(\tau),
$$
{where argmax denotes the argument that makes the function maximum.}
This maximization is performed in our algorithm through the downhill simplex method, 
restricted here to one dimension \cite{Press}.
The computation of the zeta function uses an algorithm based upon the Euler-Maclaurin 
series \cite{Vepstas},
$$
\sum_{k=0}^\infty \tilde g(k) \simeq \sum_{k=0}^{M-1} \tilde g(k) + \int_M^\infty
\tilde g(k)dk + \frac{\tilde g(M)} 2 - 
\sum_{k=1}^P \frac{B_{2k}}{(2k)!}\tilde g^{(2k-1)}(M),
$$
where $B_{2k}$ are the Bernoulli numbers 
($B_2= 1/6, B_4 = -1/30, B_6 = 1/42, B_8 = -1/30,\dots $) \cite{Abramowitz}.
The desired approximation is obtained by applying the formula
to $\tilde g(k)=(a+k)^{-\tau}$, for which the derivatives of order $2k-1$ are
$$
\tilde g^{(2k-1)}(M)=\tilde g^{(2k-3)}(M) \frac {(\tau+2k-3)(\tau+2k-2)}{(a+M)^2},
$$
with $\tilde g^{(1)}(M)=\tilde g'(M)=-\tau /(a+M)^{\tau+1}$
and the integral yields $(a+M)^{1-\tau}/(\tau-1)$.
So,
$$
\zeta(\tau,a)
\simeq \sum_{k=0}^{M-1} \frac 1 {(a+k)^\tau}  
+\frac{(a+M)^{-(\tau-1)}}{\tau-1}
+\frac 1 {2(a+M)^\tau}
 +
\sum_{k=1}^P {B_{2k}}C_{2k-1},
$$
with
$$
C_{2k-1}=\frac {(\tau+2k-2)(\tau+2k-3)}{2k(2k-1)(a+M)^2} C_{2k-3}
\, \mbox{ and } \, 
C_{1}=\frac{\tau}{2(a+M)^{\tau+1}}.
$$
The sum from $k=1$ to $P$ is stopped when a minimum value term is
reached \cite{Vepstas}, or when $k=P=18$.
We also take $M=14$. As a check, the reader can verify that
this method allows to calculate $\zeta(2,1)=\pi^2/6$ with more than
16 correct significant figures.


\subsubsection*{Kolmogorov-Smirnov goodness-of-fit test}

As, given $a$, the fit only depends on the geometric mean of the data, 
maximum likelihood can yield very bad fits if the data are not power-law
distributed
(because the estimation assumes a priori that that hypothesis holds).
%
The second step (step 2) of the procedure is to measure the deviation between 
the data and the fit. For that purpose, we use the Kolmogorov-Smirnov
statistic $d_{emp}$, defined as the maximum absolute difference between the 
(complementary) cumulative distributions
corresponding to the empirical data and to the fit 
(parameterized by $\tau=\tau_{emp}$)
\cite{Press}, 
i.e.,
$$
d_{emp}=\mbox{max}_{\forall x\ge a}\left|\frac{N_x}{N_a}-G(x)\right|,
$$
where 
the maximization is performed for all values of $x \ge a$,
integer and not integer, 
and
$N_x$ counts the number of data with values equal to or above $x$
(defined only for $x\ge a$).
Therefore, large and small values of $d_{emp}$ denote respectively bad and good fits.
We recall that although $g(x)$ is a pure power law above $a$,
$G(x)$ is only a power law asymptotically.

\subsubsection*{Simulation procedure}

The next step (step $3$) consists in the evaluation of which is good and which is bad;
this is done with simulated data 
following the distribution obtained by maximum likelihood estimation, 
that is, a discrete power law defined for $x\ge a$ with exponent $\tau_{emp}$.
The simulation procedure generalizes to the case $a > 1$ the rejection method explained 
by Devroye \cite{Devroye}. 
Although more efficient procedures have been proposed \cite{Hormann},
we were not aware of them at the time of writing and running our code.

The generalization of the method of Ref. \cite{Devroye} proceeds as follows:
First, a uniform random number $u$ is generated between 
0 and $u_{max}$, 
where $u_{max}$ 
fulfills
$a=1/u_{max}^{1/(\tau_{emp}-1)}$.
Then, a new random number $m$ is obtained as the integer part of 
$y=1/u^{1/(\tau_{emp}-1)}$,
i.e., 
$$
m=\mbox{int}(1/u^{1/(\tau_{emp}-1)}),
$$
which will verify $m \ge a$ if $u_{max}$ fulfills
$a=1/u_{max}^{1/(\tau_{emp}-1)}$.
Notice that $y$ has probability density $(\tau_{emp}-1) a^{\tau_{emp}-1}/y^{\tau_{emp}}$,
which is the continuous approximation of our $g(x)$,
and its cumulative distribution is $(a/y)^{\tau_{emp}-1}$.
The same cumulative distribution holds for $m$
(but only for its integer values).
From here, the probability mass function of $m$ turns out to be
$q(m)=(a/m)^{\tau_{emp}-1}-(a/(m+1))^{\tau_{emp}-1}$
(which, for large $m$ is $(\tau_{emp}-1) a^{\tau_{emp}-1}/m^{\tau_{emp}}$).
Then, the rejection method gives a random number $x=m$ distributed following $g(x)$
if $m$ is kept when a new uniform random number $v$ (between 0 and 1)
fulfills 
$$
v \le \frac{g(m)}{c q(m)}
$$ 
(and rejected otherwise),
with $c$ the rejection constant 
$$
c=\mbox{max}_{\forall m \ge a} \frac{g(m)}{q(m)}=
\frac{g(a)}{q(a)}=
\frac 1 {\zeta(\tau_{emp},a)a^{\tau_{emp}}}
\left(
\frac{(a+1)^{\tau_{emp}-1}}  
{(a+1)^{\tau_{emp}-1}-a^{\tau_{emp}-1}}.
\right) 
$$
The resulting random variable $x=m$ corresponds to the hidden rank $z$
in Subsec. 3.2 and to the size $n$ in 3.3.
The maximum value of $g(m)/q(m)$ takes place at $m=a$
because this is a decreasing function;
this is shown in the next subsection.
Defining the auxiliary variable $ T =(1+m^{-1})^{\tau_{emp}-1}$
and the constant $b=(a+1)^{\tau_{emp}-1}$
the acceptation condition is equivalent to
$$
v m \frac { T -1}{b-a^{\tau_{emp}-1}} \le \frac {a T} b,
$$
which shows clearly that the simulation procedure does not require the 
computation of the Hurwitz zeta function.

Once $N_a$ simulated values of $x$ have been obtained, 
they are treated in exactly the same way as the empirical data, 
following steps 1 and 2 above
\cite{Clauset}
(see also the Supporting Information of Ref. \cite{Malgrem}):
first (step 4), maximum likelihood estimation
leads to a value of the exponent, this time denoted as $\tau_{sim}$
(notice that this value will be close but distinct to $\tau_{emp}$,
{due to statistical fluctuations});
second (step 5), the Kolmogorov-Smirnov statistic leads to a 
quantification of the distance between the simulated data (with $\tau_{emp}$) 
and their fit (parameterized by $\tau_{sim}$), 
now called $d_{sim}$.

However, comparison of this single value with the empirical one, $d_{emp}$,
does not allow to draw any conclusion. 
Naturally, one needs an ensemble of values of $d_{sim}$, which are 
obtained repeating the simulation procedure (steps 3, 4, and 5) many times.
The position of $d_{emp}$ in relation to the distribution defined
by the obtained values of $d_{sim}$ allows us to calculate the $p-$value of the
fit.
This is just defined as the probability that true power-law data
(as the simulated data is), 
with exponent $\tau_{emp}$, yield a Kolmogorov-Smirnov statistic equal 
or larger than $d_{emp}$
{
(in other words, the probability that for real power-law data the fit is worse
than for the empirical data).
}
Its estimation is just
$$
p=\frac {\mbox{number of simulations with } d_{sim}\ge d_{emp}} 
{\mbox{number of simulations}}.
$$
The uncertainty in $p$ can be obtained from the fact that 
the number of simulations with  $d_{sim}\ge d_{emp}$ will be
binomially distributed, 
therefore, its standard deviation can be estimated as
$$
\sigma_p =\sqrt{\frac{p(1-p)}{\mbox{number of simulations}}}.
$$
For 1000 simulations and $p$ around 0.5 this is 0.016,
but if $p$ or $1-p$ are about 0.01 we find 0.003.
The simulation procedure also allows one to obtain the uncertainty of $\tau_{emp}$
as the standard deviation of the values of $\tau_{sim}$.

So, for a fixed value of $a$ we end with a value of $p$ that tells us
the goodness of the fit. Usually, values of $p$ below 0.05 are considered bad,
and therefore the hypothesis under testing (the goodness of the fit) 
is rejected, although this value is rather arbitrary.
Repeating the whole procedure for different values of $a$ we will obtain (or not)
a set of acceptable $a-$values, together with their corresponding exponents.
In order to select one of them, we just choose the smallest $a-$value (which yields
the largest range) provided that $p> 0.20$.
This concludes the fitting and testing procedure, leading to final values $a^*$ and $\tau^*$
(denoted in the main text simply as 
$a$ and $\tau$,
$r_a$ and $\alpha$,
or
$n_a$ and $\gamma$).
{
In a formula,
$$
a^*=\min\{a \mbox{ such that } p > 0.20 \},
$$
which has associated the resulting exponent $\tau_{emp}^*$.
It is worth mentioning that the final $p-$value of the fit for varying $a$ is not the
one corresponding to fixed $a=a^*$;
nevertheless, for our purposes its computation is not necessary.
}

\subsection*{Calculation of the rejection constant}

The efficiency of the simulations of discrete power-law distributed numbers
depends on finding the optimum rejection constant,
which is given by the maximum of $g(m)/q(m)$,
where the functions are defined in the previous subsection.
We will show that the maximum is reached for the smallest value of $m$,
as $g(m)/q(m)$ is a monotonically decreasing function.
Let us calculate (removing irrelevant multiplicative constants),
$$
\frac{q(m)}{g(m)} \propto m-\frac{m}{(1+m^{-1})^{\tau_{emp}-1}},
$$
whose derivative is
$$
\left(\frac{q(m)}{g(m)}\right)' \propto
1-\left(1+\frac 1 m\right)^{-(\tau_{emp}-1)} \left(1+\frac{\tau_{emp}-1}{m+1}\right).
$$
Then, it is enough to show that $(1+m^{-1})^{\tau_{emp}-1}$ is greater than
$1+(\tau_{emp}-1)/(m+1)$,
as this implies that the previous derivative is positive and $q(m)/g(m)$ is
monotonically increasing.
For $\tau_{emp} > 2$ we can write $1+(\tau_{emp}-1)/(m+1) < 1+(\tau_{emp}-1)m^{-1}$, 
which is indeed smaller than $(1+m^{-1})^{\tau_{emp}-1}$,
as the Bernoulli's inequality states for $\tau_{emp} >2$ and $m^{-1} > 0$.
Indeed, a version of Bernoulli's inequality states that
$1+sz < (1+z)^s$ with $s$ and $z$ any real numbers 
fulfilling $s>1$ and $z>0$.
For $1 < \tau_{emp} \le 2$ we write
$$
1+\frac{\tau_{emp}-1}{m+1}=\frac{1+(\tau_{emp})m^{-1}}{1+m^{-1}},
$$
which is again smaller than $(1+m^{-1})^{\tau_{emp}-1}$,
using the Bernoulli's inequality for $\tau_{emp} >1$
(this demonstration also holds for $\tau_{emp} > 2$,
but the previous one is simpler).

\subsection*{Logarithmic binning in the discrete case}

In the plots, the fits are compared against the empirical 
probability mass functions.
These are estimated adapting logarithmic binning 
\cite{Hergarten_book,Christensen_Moloney}, 
which uses a constant number of bins per {order of magnitude} (5 in our case), 
to discrete distributions \cite{Christensen_Moloney}.
Let us consider the intervals 
$[x_{(k)}, x_{(k+1)})$, labeled by $k=0,1\dots$ with
$x_{(k+1)}=B x_{(k)}$ and $B=\sqrt[5]{10}$ (in our case);
the starting value $x_{(0)}$ is irrelevant,
but the values of $x_{(k)}$ should not be integer,
for numerical convenience.
Then,
each occurrence of $x$ in the dataset is associated to a value of 
$k$ using the formula 
$$
k=\log_B(x/x_{(0)}).
$$
Next,
the number of occurrences of $x$ in the interval $k$ 
(i.e., the number of types with frequency in the interval range,
see the denominator of the formula below)
is divided by the total number of occurrences of any value of $x$
(which is $N(a)=N_a$, changing notation for convenience) 
and by $\mbox{int}(x_{(k+1)})-\mbox{int}(x_{(k)})$
(which counts the number of possible values of $x$ in the interval,
i.e., the number of integers).
This yields the estimated value of the probability mass function 
$g_{emp}(x^*_{k})$
in the $k-$th interval,
$$
g_{emp}(x^*_{k})=\frac 
{N(\mbox{int}(x_{(k)})+1)-N(\mbox{int}(x_{(k+1)})+1)}
{N(a) [\mbox{int}(x_{(k+1)})-\mbox{int}(x_{(k)})]},
$$
where the value of the probability mass function is associated to a point $x^*_{k}$
in the interval given by the geometric mean of the smallest and largest integer
in the interval,
$$
x^*_{k}=\sqrt{\mbox{int}(x_{(k)}+1)\mbox{int}(x_{(k+1)})},
$$
see Ref. \cite{Hergarten_book}.
{Compare the last two formulas with Eq. 1.12 in Ref. \cite{Baayen}.
Notice that our procedure estimates directly the probability mass 
function for small values of $x$ (as the number of integers in each bin is one, or zero),
but tends to its continuous version (the probability density)
for large $x$ (as the number of integers approaches the width of the bin).}
Estimation of probability distributions for discrete
but non-integer variables was described in Ref. \cite{Deluca_npg}.

{
The error bars associated to $g_{emp}(x)$
can be estimated from the fact that the number of counts
in a given bin can be considered binomially distributed 
(assuming that the data are not correlated, 
but this assumption is also made in order to apply maximum likelihood
estimation; in practice it is enough that the number of data 
is much larger than the range of correlations).
For a binomial variable the ratio between the standard deviation 
and the mean (the relative error) is given, approximately, 
by the inverse of the square root of the mean number of counts
(if there is no bin that accumulates most of the counts).
The same relation holds for $g_{emp}(x)$, because it is proportional 
to the number of counts and the proportionality constants
vanish when the ratio standard deviation $/$ mean is taken.
Approximating the mean number of counts to the actual number of counts, then,
the standard deviation of $g_{emp}(x)$ in bin $k$ is obtained as
$$
\sigma_k \simeq \frac {g_{emp}(x_k^*)}{\sqrt{\mbox{counts in } k}} =
\frac {g_{emp}(x_k^*)}{\sqrt{N(\mbox{int}(x_{(k)})+1)-N(\mbox{int}(x_{(k+1)})+1)}}.
$$
}

Finally,
notice that the estimation of the empirical mass function does not play any role
in the fitting and testing procedures, and it is shown in the plots
just for illustrative purposes.
There, we compare $g_{emp}(x)$, defined for $x\ge 1$,
with the fit $g(x)$ defined for $x\ge a$; then, a correction 
constant is applied to the latter in order that the
fit is properly displayed.
So, $g_{emp}(x)$ is plotted together with $g(x)N_a/N$.
In the case of the rank-size representation
this means that we plot the empirical $n_{emp}(r)/L_{tot}$
together with the theoretical $(n(r)/L)(L/L_{tot})=n(r)/L_{tot}$.


\begin{thebibliography}{10}

\bibitem{Bak_book}
P.~Bak.
\newblock {\em How Nature Works: The Science of Self-Organized Criticality}.
\newblock Copernicus, New York, 1996.

\bibitem{Sornette_critical_book}
D.~Sornette.
\newblock {\em Critical Phenomena in Natural Sciences}.
\newblock Springer, Berlin, 2nd edition, 2004.

\bibitem{Mitz}
M.~Mitzenmacher.
\newblock A brief history of generative models for power law and lognormal
  distributions.
\newblock {\em Internet Math.}, 1 (2):226--251, 2004.

\bibitem{Newman_05}
M.~E.~J. Newman.
\newblock Power laws, {Pareto} distributions and {Zipf}'s law.
\newblock {\em Cont. Phys.}, 46:323 --351, 2005.

\bibitem{Simkin11}
M.V. Simkin and V.P. Roychowdhury.
\newblock Re-inventing {Willis}.
\newblock {\em Physics Reports}, 502(1):1 -- 35, 2011.

\bibitem{Bauke}
H.~Bauke.
\newblock Parameter estimation for power-law distributions by maximum
  likelihood methods.
\newblock {\em Eur. Phys. J. B}, 58:167--173, 2007.

\bibitem{White}
E.~P. White, B.~J. Enquist, and J.~L. Green.
\newblock On estimating the exponent of power-law frequency distributions.
\newblock {\em Ecol.}, 89:905--912, 2008.

\bibitem{Clauset}
A.~Clauset, C.~R. Shalizi, and M.~E.~J. Newman.
\newblock Power-law distributions in empirical data.
\newblock {\em SIAM Rev.}, 51:661--703, 2009.

\bibitem{Corral_nuclear}
A.~Corral, F.~Font, and J.~Camacho.
\newblock Non-characteristic half-lives in radioactive decay.
\newblock {\em Phys. Rev. E}, 83:066103, 2011.

\bibitem{Corral_Deluca}
A.~Deluca and A.~Corral.
\newblock Fitting and goodness-of-fit test of non-truncated and truncated
  power-law distributions.
\newblock {\em Acta Geophys.}, 61:1351--1394, 2013.

\bibitem{Corral_Gonzalez}
A.~Corral and A.~Gonz\'alez.
\newblock Power-law distributions in geoscience revisited.
\newblock {\em Earth Space Sci.}, submitted:submitted, 2018.

\bibitem{Voitalov_krioukov}
Ivan {Voitalov}, Pim {van der Hoorn}, Remco {van der Hofstad}, and Dmitri
  {Krioukov}.
\newblock {Scale-free Networks Well Done}.
\newblock {\em arXiv}, 1811.02071, 2018.

\bibitem{Zipf1949a}
G.~K. Zipf.
\newblock {\em Human behaviour and the principle of least effort}.
\newblock Addison-Wesley, Cambridge (MA), USA, 1949.

\bibitem{Adamic_Huberman}
L.~A. Adamic and B.~A. Huberman.
\newblock {Zipf's} law and the {Internet}.
\newblock {\em Glottom.}, 3:143--150, 2002.

\bibitem{Furusawa2003}
C.~Furusawa and K.~Kaneko.
\newblock Zipf's law in gene expression.
\newblock {\em Phys. Rev. Lett.}, 90:088102, 2003.

\bibitem{Axtell}
R.~L. Axtell.
\newblock Zipf distribution of {U.S.} firm sizes.
\newblock {\em Science}, 293:1818--1820, 2001.

\bibitem{Serra_scirep}
J.~Serr\`a, A.~Corral, M.~Bogu{\~n}\'a, M.~Haro, and J.~Ll. Arcos.
\newblock Measuring the evolution of contemporary western popular music.
\newblock {\em Sci. Rep.}, 2:521, 2012.

\bibitem{Baayen}
H.~Baayen.
\newblock {\em Word Frequency Distributions}.
\newblock Kluwer, Dordrecht, 2001.

\bibitem{Baroni2009}
M.~Baroni.
\newblock Distributions in text.
\newblock In A.~L\"udeling and M.~Kyt\"o, editors, {\em Corpus linguistics: An
  international handbook, Volume 2}, pages 803--821. Mouton de Gruyter, Berlin,
  2009.

\bibitem{Piantadosi}
S.~T. Piantadosi.
\newblock Zipf's law in natural language: a critical review and future
  directions.
\newblock {\em Psychon. Bull. Rev.}, 21:1112--1130, 2014.

\bibitem{Malevergne_Sornette_umpu}
Y.~Malevergne, V.~Pisarenko, and D.~Sornette.
\newblock Testing the {Pareto} against the lognormal distributions with the
  uniformly most powerful unbiased test applied to the distribution of cities.
\newblock {\em Phys. Rev. E}, 83:036111, 2011.

\bibitem{Lu_2010}
L.~L\"u, Z.-K. Zhang, and T.~Zhou.
\newblock Zipf's law leads to {Heaps}' law: Analyzing their relation in
  finite-size systems.
\newblock {\em PLoS ONE}, 5(12):e14139, 12 2010.

\bibitem{Zanette2012a}
D.~Zanette.
\newblock {\em Statistical patterns in written human language}.
\newblock 2012.

\bibitem{Mandelbrot61}
B.~Mandelbrot.
\newblock {On the theory of word frequencies and on related {Markovian} models
  of discourse}.
\newblock In R.~Jakobson, editor, {\em Structure of Language and its
  Mathematical Aspects}, pages 190--219. American Mathematical Society,
  Providence, RI, 1961.

\bibitem{pawitan2001}
Y.~Pawitan.
\newblock {\em In All Likelihood: Statistical Modelling and Inference Using
  Likelihood}.
\newblock Oxford UP, Oxford, 2001.

\bibitem{Gerlach_Altmann}
M.~Gerlach and E.~G. Altmann.
\newblock {Stochastic model for the vocabulary growth in natural languages}.
\newblock {\em Phys. Rev. X}, 3:021006, 2013.

\bibitem{Baixeries2012a}
J.~Baixeries, B.~Elvev{\aa}g, and R.~{Ferrer-i-Cancho}.
\newblock The evolution of the exponent of {Zipf}'s law in language ontogeny.
\newblock {\em PLoS ONE}, 8(3):e53227, 2013.

\bibitem{Tuldava1996}
J.~Tuldava.
\newblock The frequency spectrum of text and vocabulary.
\newblock {\em J. Quantitative Linguistics}, 3(1):38--50, 1996.

\bibitem{Balasubrahmanyan1996}
V.~K. Balasubrahmanyan and S.~Naranan.
\newblock Quantitative linguistics and complex system studies.
\newblock {\em J. Quantitative Linguistics}, 3(3):177--228, 1996.

\bibitem{Ferrer2000a}
R.~{Ferrer i Cancho} and R.~V. Sol\'e.
\newblock Two regimes in the frequency of words and the origin of complex
  lexicons: {Zipf's} law revisited.
\newblock {\em J. Quant. Linguist.}, 8(3):165--173, 2001.

\bibitem{Petersen_scirep}
A.~M. Petersen, J.~N. Tenenbaum, S.~Havlin, H.~E. Stanley, and M.~Perc.
\newblock Languages cool as they expand: Allometric scaling and the decreasing
  need for new words.
\newblock {\em Sci. Rep.}, 2:943, 2012.

\bibitem{Ferrer2009a}
R.~{Ferrer-i-Cancho} and R.~Gavald\`a.
\newblock The frequency spectrum of finite samples from the intermittent
  silence process.
\newblock {\em Journal of the American Association for Information Science and
  Technology}, 60(4):837--843, 2009.

\bibitem{Altmann_Gerlach}
E.~G. Altmann and M.~Gerlach.
\newblock Statistical laws in linguistics.
\newblock In M.~D. Esposti, E.~G. Altmann, and F.~Pachet, editors, {\em
  Creativity and Universality in Language. {Lecture Notes in Morphogenesis}}.
  Springer, 2016.

\bibitem{FontClos_Corral}
F.~Font-Clos, G.~Boleda, and A.~Corral.
\newblock A scaling law beyond {Zipf}'s law and its relation with {Heaps}' law.
\newblock {\em New J. Phys.}, 15:093033, 2013.

\bibitem{Heaps_1978}
H.~S. Heaps.
\newblock {\em Information retrieval: computational and theoretical aspects}.
\newblock Academic Press, 1978.

\bibitem{Abramowitz}
M.~Abramowitz and I.~A. Stegun, editors.
\newblock {\em Handbook of Mathematical Functions}.
\newblock Dover, New York, 1965.

\bibitem{Johnson_univariate}
N.~L. Johnson, A.~W. Kemp, and S.~Kotz.
\newblock {\em Univariate Discrete Distributions}.
\newblock Wiley-Interscience, New Jersey, 3rd edition, 2005.

\bibitem{Peters_Deluca}
O.~Peters, A.~Deluca, A.~Corral, J.~D. Neelin, and C.~E. Holloway.
\newblock Universality of rain event size distributions.
\newblock {\em J. Stat. Mech.}, P11030, 2010.

\bibitem{Goldstein}
M.~L. Goldstein, S.~A. Morris, and G.~G. Yen.
\newblock Problems with fitting to the power-law distribution.
\newblock {\em Eur. Phys. J. B}, 41:255--258, 2004.

\bibitem{Pueyo}
S.~Pueyo and R.~Jovani.
\newblock Comment on ``{A} keystone mutualism drives pattern in a power
  function''.
\newblock {\em Science}, 313:1739c--1740c, 2006.

\bibitem{Hergarten_book}
S.~Hergarten.
\newblock {\em Self-Organized Criticality in Earth Systems}.
\newblock Springer, Berlin, 2002.

\bibitem{Burroughs}
S.~M. Burroughs and S.~F. Tebbens.
\newblock Power-law scaling and probabilistic forecasting of tsunami runup
  heights.
\newblock {\em Pure Appl. Geophys.}, 162:331--342, 2005.

\bibitem{Casella}
G.~Casella and R.~L. Berger.
\newblock {\em Statistical Inference}.
\newblock Duxbury, Pacific Grove CA, 2nd edition, 2002.

\bibitem{Corral_Deluca_arxiv}
A.~Corral, A.~Deluca, and R.~{Ferrer-i-Cancho}.
\newblock A practical recipe to fit discrete power-law distributions.
\newblock {\em ArXiv}, 1209:1270, 2012.

\bibitem{Devroye}
L.~Devroye.
\newblock {\em Non-Uniform Random Variate Generation}.
\newblock Springer-Verlag, New York, 1986.

\bibitem{Christensen_Moloney}
K.~Christensen and N.~R. Moloney.
\newblock {\em Complexity and Criticality}.
\newblock Imperial College Press, London, 2005.

\bibitem{Deluca_npg}
A.~Deluca and A.~Corral.
\newblock Scale invariant events and dry spells for medium-resolution local
  rain data.
\newblock {\em Nonlinear Proc. Geophys.}, 21:555--567, 2014.

\bibitem{Kolmogorov1956aa}
A.~N. Kolmogorov.
\newblock {\em Foundations of the Theory of Probability}.
\newblock Chelsea Publising Company, New York, 2nd edition, 1956.

\bibitem{Bouchaud_georges}
J.-P. Bouchaud and A.~Georges.
\newblock Anomalous diffusion in disordered media: statistical mechanisms,
  models and physical applications.
\newblock {\em Phys. Rep.}, 195:127--293, 1990.

\bibitem{Corral_csf}
A.~Corral.
\newblock Scaling in the timing of extreme events.
\newblock {\em Chaos. Solit. Fract.}, 74:99--112, 2015.

\bibitem{Corral_Font_Clos_PRE17}
A.~Corral and F.~Font-Clos.
\newblock Dependence of exponents on text length versus finite-size scaling for
  word-frequency distributions.
\newblock {\em Phys. Rev. E}, 96:022318, 2017.

\bibitem{Burnham2002a}
K.~P. Burnham and D.~R. Anderson.
\newblock {\em Model selection and multimodel inference. A practical
  information-theoretic approach}.
\newblock Springer, New York, 2nd edition, 2002.

\bibitem{Ferrer2004a}
R.~{Ferrer i Cancho}.
\newblock The variation of {Zipf's} law in human language.
\newblock {\em Eur. Phys. J. B}, 44:249--257, 2005.

\bibitem{Press}
W.~H. Press, S.~A. Teukolsky, W.~T. Vetterling, and B.~P. Flannery.
\newblock {\em Numerical Recipes in {FORTRAN}}.
\newblock Cambridge University Press, Cambridge, 2nd edition, 1992.

\bibitem{Vepstas}
L.~Vepstas.
\newblock An efficient algorithm for accelerating the convergence of
  oscillatory series, useful for computing the polylogarithm and {Hurwitz} zeta
  functions.
\newblock {\em ArXiv}, page math/0702243, 2007.

\bibitem{Hormann}
W.~H\"{o}rmann and G.~Derflinger.
\newblock Rejection-inversion to generate variates from monotone discrete
  distributions.
\newblock {\em ACM Trans. Model. Comput. Simul.}

\bibitem{Malgrem}
R.~D. Malmgren, D.~B. Stouffer, A.~E. Motter, and L.~A.~N. Amaral.
\newblock A {Poissonian} explanation for heavy tails in e-mail communication.
\newblock {\em Proc. Natl. Acad. Sci. USA}, 105:18153--18158, 2008.

\end{thebibliography}

\end{document}